\documentclass{aa}  

\usepackage[pdfpagelabels=false]{hyperref}	
\hypersetup{colorlinks=true,linkcolor=blue,citecolor=blue,filecolor=blue,urlcolor=blue,}
\usepackage{natbib}
\usepackage{graphicx}
\usepackage{xcolor}
\usepackage{txfonts}
\usepackage{lipsum}
\usepackage{subcaption}  
\usepackage{lscape}             
\usepackage{placeins}           

\usepackage{booktabs}

\newcommand{\msun}{${\rm M}_{\odot}$ }
\newcommand{\msunperyr}{${\rm M}_{\odot}$ yr$^{-1}$ }
\newcommand{\dexre}{dex r$_{50}^{-1}$ }

\newcommand{\reff}{$r_{50}$ }

\newcommand{\ropt}{$r_{83}$ }
\newcommand{\rvir}{$r_{\rm vir}$ }
\newcommand{\grad}{\hbox{$\nabla_{{\rm (O/H)}}$} }
\newcommand{\cielo}{{\sc CIELO}}
\newcommand{\innergrad}{$\nabla^{\rm Inner}_{\text{\scriptsize(O/H)}}$ }
\newcommand{\midgrad}{$\nabla^{\rm Mid}_{\text{\scriptsize(O/H)}}$ } 
\newcommand{\outergrad}{$\nabla^{\rm Outer}_{\text{\scriptsize(O/H)}}$ }
\newcommand{\dbagrad}{$\nabla_{\text{\scriptsize(O/H)}}$ } 
\newcommand{\tsgrad}{$\nabla^{\rm TS}_{\text{\scriptsize(O/H)}}$ } 
\newcommand{\accrates}{$\Dot{\mathcal{M}}$ }

\defcitealias{Belfiore2017}{B17}
\defcitealias{SanchezM2018}{SM18}
\defcitealias{Tissera2019}{T19}
\defcitealias{Tissera2021}{T22}

\renewcommand\makeLineNumber{}

\begin{document}

   \title{Insight into the physical processes that shape the metallicity profiles in galaxies}

   \author{Brian Tapia-Contreras\inst{1, 2}\thanks{E-mail: \url{brian.tapia@uc.cl}} \and 
           Patricia B. Tissera\inst{1, 2, 3}\and
           Emanuel Sillero\inst{1, 3}\and
           Jenny Gonzalez-Jara\inst{1, 2}\and
           Catalina Casanueva-Villarreal\inst{1, 2}\and
           Susana Pedrosa\inst{4}\and
           Lucas Bignone\inst{4}\and
           Nelson D. Padilla\inst{5, 6}\and
           Rosa Domínguez-Tenreiro\inst{7}
        }

   \institute{Instituto de Astrofísica, Pontificia Universidad Católica de Chile, Av. Vicuña Mackenna 4860, Santiago, Chile. \and 
    Centro de Astro-ingeniería, Pontificia Universidad Católica de Chile, Av. Vicuña Mackenna 4860, Santiago, Chile.\and
    Millenium Nucleus ERIS\and
    Instituto de Astronomía y Física del Espacio, CONICET-UBA, Casilla de Correos 67, Suc. 28, 1428, Buenos Aires, Argentina.\and
    Instituto de Astronomía Teórica y Experimental (IATE, CONICET-Universidad Nacional de Córdoba), Laprida 854, X5000BGR, Córdoba, Argentina\and
    Observatorio Astronómico de la Universidada Nacional de Córdoba, Laprida 854, X5000BGR, Córdoba, Argentina\and
    Departamento de Física Teórica, Universidad Autónoma de Madrid, E-28049 Cantoblanco, Madrid, Spain
    }

   \date{Received September 30, 20XX}

 
  \abstract
   {The distribution of chemical elements in the star-forming regions can store information on the chemical enrichment history of the galaxies, and particularly of recent events. Negative metallicity gradients are expected in galaxies forming inside-out. Azimuthal-averaged profiles are usually fitted to the projected chemical distributions to quantify them. However, observations show that the metallicity profiles can be broken. }
   {We aim to study the diversity of metallicity profiles that can arise in the current cosmological context and compare them with available observations. Additionally, we seek to identify the physical processes responsible for breaks in metallicity profiles by using two galaxies as case studies.} 
   {We analyze central galaxies from the cosmological simulations of \cielo~project, with stellar masses within the range of $10^{8.5}$ to $10^{10.5 }$ \msun at $z=0$. A new algorithm, DB-A, was developed to fit multiple power laws to the metallicity profiles, enabling a flexible assessment of metallicity gradients in various galactic regions. The simulations include detailed modeling of gas components, metal-dependent cooling, star formation, and supernova feedback. }
   {At $z=0$, we find a diversity of shapes, with inner and outer drops and rises, with a few galaxies having double breaks. Inner, outer and middle gradients agree with observations. We also find that using a single linear regression to fit gradients usually traces well the middle gradient. 
    A detailed temporal analysis of the main galaxies of a Local Group analogues reveals the occurrence of inner and outer breaks at all cosmic times, with the latter ones being the most common feature during the evolution of our case studies. Significant variability in the metallicity gradients is found at high redshift, transitioning to more gradual evolution at lower redshifts. Most of inner breaks have enhanced oxygen abundances in the center, linked to gas accretion followed by  efficient star formation. Inner breaks with diluted oxygen abundances are less common and are found in galaxies with disrupted gas distributions, affected by feedback-driven ejection of enriched gas.  Outer breaks with high abundances are linked to processes such as the re-accretion of enriched material, extended star formation, and enhanced gas mixing from the CGM. Outer breaks with diluted metallicities in the outskirts are found mainly at high redshift and are associated with the accretion of metal-poor gas from cold flows. We also highlight and illustrate the complex interplay of these processes which act often together.  }
   {}


   \maketitle

\section{Introduction}

The chemical abundance of star-forming regions and stellar populations store relevant information about the history of galaxy formation \citep{Tinsley1980, MaiolinoManucci, jaraferreira2024}. Most of heavy chemical elements are produced by stellar nucleosynthesis and injected into the interstellar medium (ISM) by supernovae (SNe) explosions and stellar winds  \citep{Matteuccibook, Naab2017}. In the current cosmological paradigm, the Lambda Cold Dark Matter ($\Lambda$CDM) scenario, the structure is  assembled hierarchically, with the gas cooling and collapsing within  dark matter halos, which grow by accretion and mergers.
Within  halos, the cool gas settles and, eventually, star formation activity  takes place \citep{White&Rees1978}.

The formation of disc galaxies can be explained by the so-called inside-out formation model with overall angular momentum conservation \citep{FallEfstathiou}, in which the central regions formed first. The global conservation of angular momentum results from a balance between the infalling  and the expelled material \citep[e.g.][]{Sales2012,Pedrosa2015,somervilledave2015}. Within this scenario, the central regions of galaxies are enriched before galaxy outskirts and for a longer time, resulting in a negative metallicity gradient. Observations of HII regions in disc galaxies have detected  since the 1970s \citep{Searle1971, Peimbert1979}. Recent integral field units (IFU) observations, such as MaNGA \citep{Bundy2015_MaNGA}, CALIFA \citep{Garcia2015_CALIFA} and MUSE-Wide \citep{Urrutia2019_MUSE}, have provided high quality data useful to study the spatial distribution of metallicity in galaxies in star-forming regions \citep[e.g.][]{Sanchez2014, SanchezM2016}.

Observational results showed the existence of mass (luminosity)-metallicity gradient relation, MGMR, in the Local Universe \citep{vilacostas1992, zaritsky1994, garnett1997}. However, with the advent of large database, the MGMR has been reshaped. \citet{Belfiore2017} found a dependence of the gas-phase metallicity gradient on stellar mass in non-interacting nearby galaxies from the MaNGA survey, suggesting that low-mass galaxies (log(M$_\star$/M$_\odot)\sim9.5$) and high-mass galaxies (log(M$_\star$/M$_\odot)\sim11$) exhibit shallower gradients than intermediate-mass galaxies with log(M$_\star$/M$_\odot)\sim10.3$ \citep[see also][]{khorambelfiore2025}. Flat gradients are expected in high-mass galaxies as the result of their more advanced evolutionary stage \citep{Molla2019}. Galactic winds have been also proposed to contribute that behavior at the low mass end, since their effects are greater in galaxies with shallow potential wells \citep{DekelSilk1986}, allowing them to more easily remove metal-rich material from the inner regions of the galaxy. They also found a systematic flattening in the center of the most massive galaxies and in the outskirts of some galaxies (even an inversion of the gradient in the later). According to the baryon cycle \citep[i.e. the interplay between gas inflows, outflows and star formation,][]{Peroux20}, this could be consistent with an scenario where metal-rich gas is ejected from the inner regions, feeds a metal-rich gas halo and simultaneously the outer regions of the galaxy re-accrete this gas \citep[e.g.][]{diaz1989,perez2011}. If galaxies are in interactions, the transfer of material between them and the opening of the spiral arms in response to the tidal torque generated during the interaction could also contribute to change the metallicity gradients in the outskirts \citep{sillero2017}. 

Observations also report  significant deviations from a steep negative gradient  in interacting galaxies and mergers compared to isolated ones \citep{Kewley2010}. Inverted or positive gradients, i.e. less enriched central regions than the outer ones, were also detected with a increasing frequency with increasing redshift \citep{Troncoso2014}. However, new observations show a large diversity of metallicity gradients even up to $z \sim 8 $ \citep{venturi2024}.  Several numerical works have reproduced shallower and inverted metallicity gradients as a result of mergers, interactions or even strong energetic feedback \citep[e.g.][]{rupke2010a, Tissera2019}.

In terms of the evolution of the gas metallicity gradients, there are several studies using different simulations and subgrid models which predict different level of evolution mainly depending on the strength of feedback \citep{Pilkington2012, Gibson2013, stinson2013, Tissera2016, ma2017, ibrahim2025, garcia2025}. Recently \citet[][\citetalias{Tissera2021}]{Tissera2021}, analysing simulated star-forming galaxies, with M$_\star \ge 10^9$ M$_\odot$, from the EAGLE project in a redshift range $z = [0.0, 2.5]$, found that the median oxygen abundance gradient is close to zero at all $z$, but exhibits an increase in the scatter at higher redshift. Such mild evolution is predicted in simulations with enhanced supernova feedback \citep[MaGICC simulation,][]{Gibson2013}. \citetalias{Tissera2021} also explored the stellar mass dependence of the metallicity gradients, finding that these gradients exhibit a weak positive dependence on stellar mass at $z \sim 0$, trend that becomes stronger with increasing $z$. Observations of high redshift galaxies using the Very Large Telescope (VLT) \citep{wuyts2016, carton2018}, Hubble Space Telescope (HST) \citep{jones2015, wang2020, simons2021, li2022} and James Webb Space Telescope (JWST) \citep{wang2022, arribas2024, venturi2024, ju2024} often report shallow metallicity gradients with significant scatter and an increasing fraction of positive metallicity gradients with redshift. Simulations are able to reproduce global these observations, however, at the very high redshift, it is difficult to model the physical processes with enough resolution.  

Most  observational and numerical  studies of metallicity gradients  assume a single linear relation to describe the radial metallicity profiles, in general, within $[0.5, 1.5] \rm r_{\rm 50}$, where $ \rm r_{\rm 50}$ is the half-mass radius. However, the presence of inner and outer breaks of the metallicity profile in certain galaxies have been reported for decades \citep[e.g.][respectively]{martinroy1995, diaz1989} and confirmed by the larger statistical results obtained with IFU observations \citep{Sanchez2014}.
\citet{SanchezM2018} \citepalias[hereafter][]{SanchezM2018} looked for deviations from single gradients by automatically fitting the metallicity profiles in nearby galaxies, using IFU observations provided by MUSE. They found that apart from the negative trend, inner drops and outer flattenings are common in disc galaxies \citep[see also][]{easeman2022,cardoso2024}. 

Numerical simulations of interacting galaxies have shown that gas inflows can be triggered  by  tidal torques generated during these events, which retrieve angular momentum and induce bar formation \citep[e.g.][]{BarnesHernquist1996,MihosHernquist1996,Tissera2016}. These gas inflows dilute the central abundances flattening the metallicity gradients \citep{perez2011,Torrey2012}. However, they can also feed new star formation and consequently, increase the level of oxygen making the metallicity profiles more negative again \citep{sillero2017, pan2025}, possibly associated to the formation of bars \citep{Chen2023}. The new born stars will produce Supernova (SN) feedback which could trigger metal-loaded outflows, transporting material out of the galaxy \citep{Pilkington2012}. Depending on the strength of the star formation activity and the potential well of the galaxies, the ejection of material might invert the metallicity gradients as shown by numerical simulations \citep{stinson2013,Tissera2019}.The metallicity gradients could recover again a weak negative metallicity gradients within around $[1.4-2]$ Gyr \citepalias{Tissera2021}.
Even if galactic outflows are not powerful enough to imprint strong change of slope of the metallicity profiles, they can weaken them and induce galactic fountains which transport the enriched material to the disc and outer parts changing the slope of the metallicity profiles \citep[e.g.][]{rupke2010b, perez2011,Torrey2012}.
Within a cosmological context,  \citet{Garcia2023} carried out a statistical study of the metallicity gradients in the Illustris-TNG50 simulation, finding a break, which separates an inner highly enriched region and  a mixing dominated outer disc region. 

In this work, we aim to characterize the shape of the metallicity gradients of simulated galaxies in a cosmological context, and analyze  the physical process participating in shaping them. For this purpose, we  present a new fitting algorithm, DB-A, to automatically classify metallicity profiles.   In this paper, we  use the suite of cosmological simulations of the Chemo-dynamIc propErties of gaLaxies and the cOsmic web project, \cielo~\citep{Tissera2025}. We characterize the metallicity profiles of the star-forming regions of 45 central galaxies with stellar masses within M$_\star $ = [$10^{8.5}, 10^{10.5}$] \msun at $z = 0$. We estimate the simulated break radius and compare them with available observations.  To study the  different physical processes acting to shape the distribution of chemical elements in the ISM, we analyze the metallicity profiles of the star-forming gas in the two main galaxies of a Local Group (LG) type environment as case studies. The simulation has adequate cadence ($\sim 0.16 $ Gyr) to follow the impact of different process on the shape of the metallicity distribution. We selected this LG analogue as a case study because it was also analyzed in two previous papers. \citet{Rodriguez2022} studied the impact of the environment on infall disc satellites, which also interacted with the central galaxies. \citet{Cataldi2023} focused on the impact of galaxy assembly on the shape of the dark matter distribution. Both works followed the infalling material and their impact. Hence, these works complement each other in relation to the study of galaxy formation and the impact of environment on baryons and dark matter.

This paper is organized as follows. Section \ref{sec:simulations} introduces the \cielo~simulations and the DB-A algorithm, used to automatically fit the metallicity profiles. In Section \ref{sec:redshift0} we report the occurrence of breaks in metallicity profiles at $z=0$, as well as the relation between the metallicity gradients and stellar mass, star formation and galaxy size. We also explore the break radius dependence on stellar mass and star formation. In Section \ref{sec:evolution} we present the evolution of the metallicity gradients of our two case studies, including the identification of star formation rates, gas accretion rates, satellite interactions and mergers. In Section \ref{sec:discussion} we discuss in detail about the physical origin of inner and outer breaks along the evolutionary history of the case studies in the context of the previous results. Finally, in Conclusions we summarize our main findings.

\section{\cielo~simulations}
\label{sec:simulations}

\begin{figure*}
 \centering
 \includegraphics[width=1.6\columnwidth]{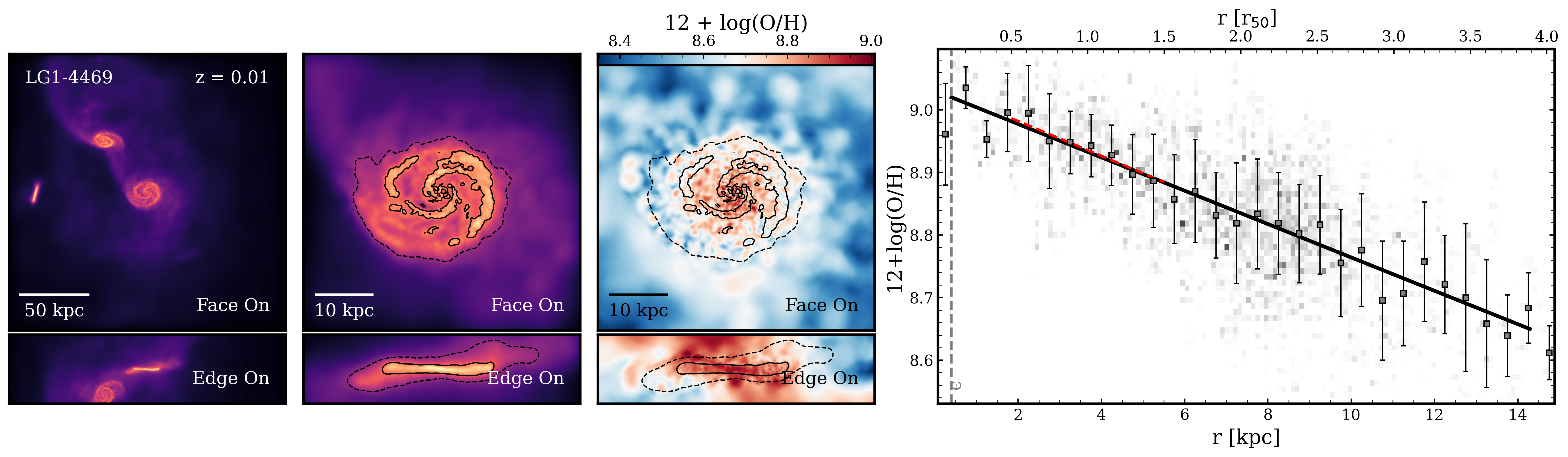}\\
 \includegraphics[width=1.6\columnwidth]{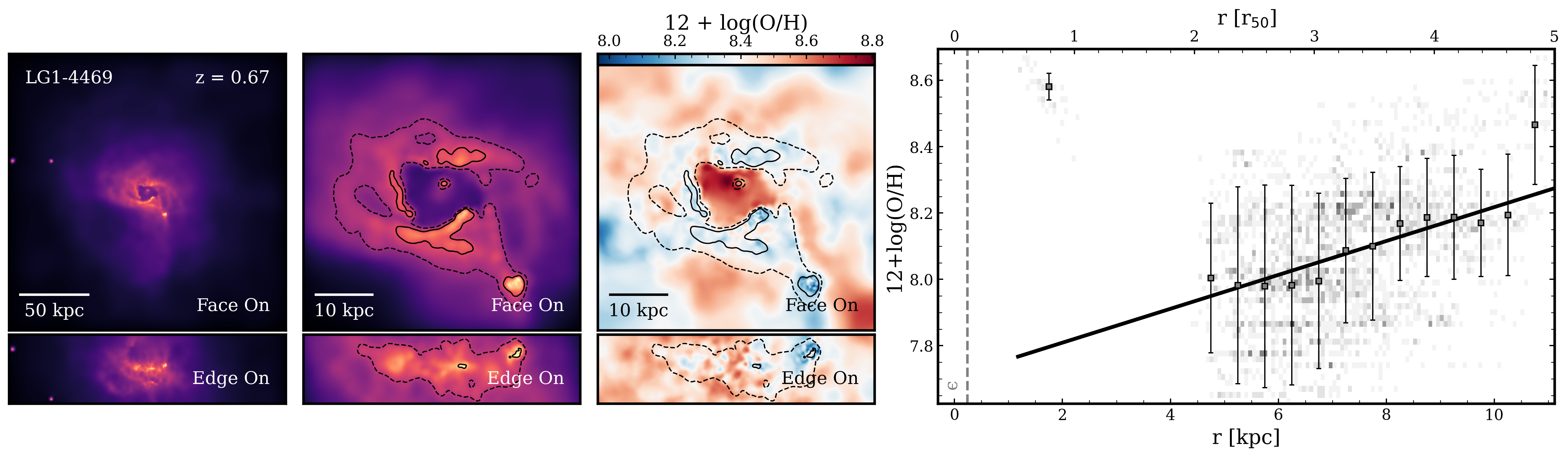}\\
 \includegraphics[width=1.6\columnwidth]{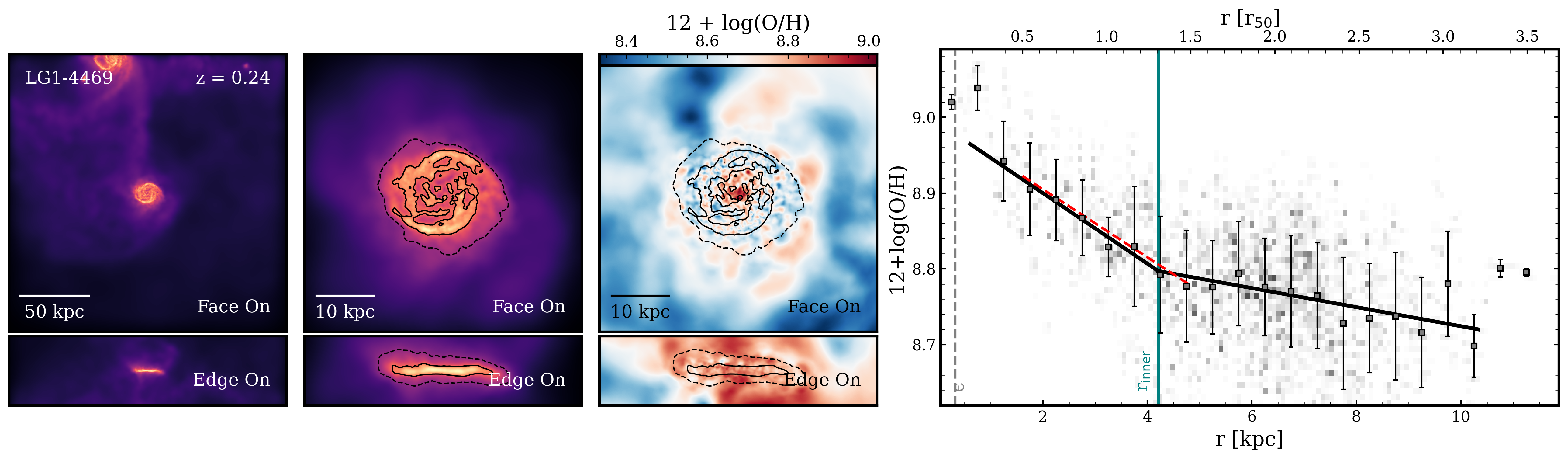}\\
 \includegraphics[width=1.6\columnwidth]{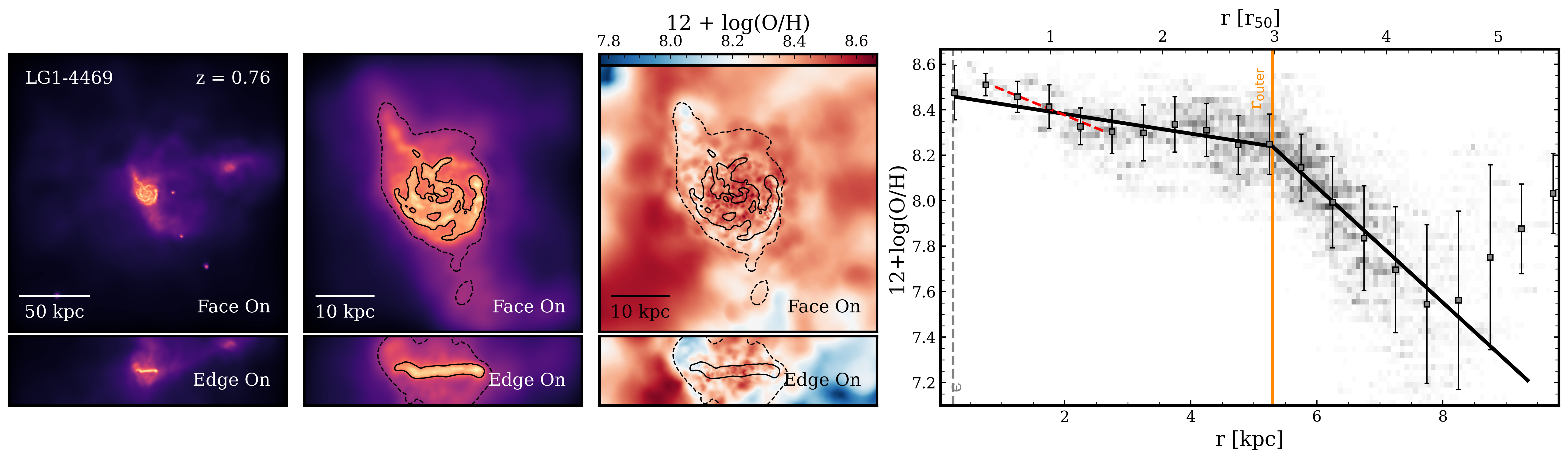}\\
 \includegraphics[width=1.6\columnwidth]{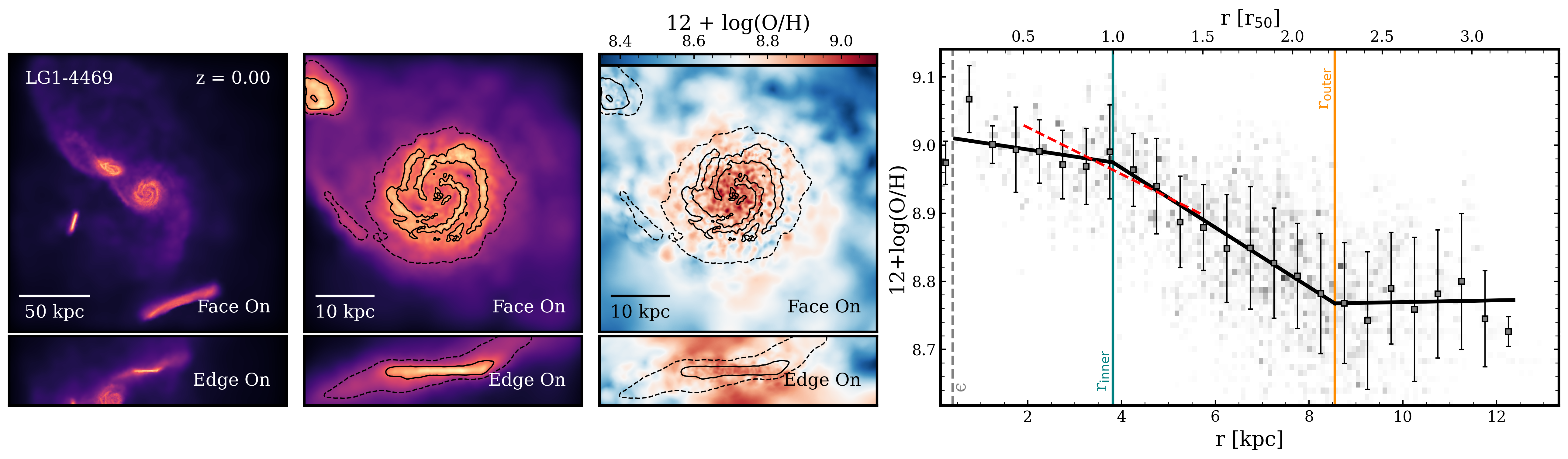}
 \caption{Examples of the variety of metallicity profiles in the \cielo~galaxies. From left to right: (i) Face-on and edge-on (top and bottom subpanels) gas density projections in a large cubic volume of 200 kpc-side centered at a central galaxy; (ii) Face-on and edge-on gas density projections in a cubic volume of 50 kpc-side located at the center of mass of the galaxy; (iii) Face-on and edge-on projected metallicity maps in the 50 kpc-side  volume. Arbitrary density contours are included in (ii) and (iii) to illustrate the morphology; (iv) Gas-phase metallicity profile of the central galaxy (background heat map). For illustrative purposes, median values within 0.5 kpc bins are shown as filled squares, with error bars indicating the standard deviation in each bin. The DB-A fit adopted in this work (see Sec. \ref{sec:constructionandfit}) is shown as a black solid line. For comparison, a linear fit within [0.5-1.5] \reff is also displayed (red dashed line). Vertical lines are included to indicate the gravitational softening $\epsilon$ (gray, dashed line), inner break radii (teal solid line) and/or outer break radii (orange solid line) when appropriate. }
 \label{fig:proj+prof}
\end{figure*}

The \cielo~project aims to study galaxy formation and the chemical evolution of baryons in different field and filamentary environments. The initial conditions (ICs) have been chosen to be centered in halos with virial masses in the range M$_{200} = [10^{10}, 10^{12}]$ \msun (Pehuen halos), and includes two LG analogues.

The \cielo~ICs are taken from a dark matter-only run of a cosmological periodic cubic box of side length $L=100$ Mpc $h^{-1}$ consistent with  a $\Lambda$CDM universe model with  $\Omega_0 = 0.3170$, $\Omega_\Lambda = 0.6825$, $\Omega_B = 0.049$ and $h = 0.6711$.  Zoom-in initial conditions were generated using the MUSIC code \citep{Hahn2011}. 
The L12 level corresponds to high resolution with $\rm m_{\rm dm}=1.36\times 10^5 h^{-1}$\msun, while the L11 level represents intermediate resolution with $\rm m_{\rm dm}=1.28\times 10^6 h^{-1}$\msun. The initial gas masses are $\rm m_{\rm gas}=2.1\times 10^4 h^{-1}$\msun~for L12 and $\rm m_{\rm gas}=2.0\times 10^5 h^{-1}$~\msun for L11. At L11 resolution, the gravitational softening values are $\epsilon_{\rm g}=400$ pc for gas and stellar particles and $\epsilon_{\rm dm}= 800$ pc for dark matter particles. In contrast, the L12 runs have $\epsilon_{\rm g}=250$ pc for gas and stellar particles and $\epsilon_{\rm dm}= 500$ pc for dark matter particles.
The detailed description of the initial conditions and the  comparison of the main properties of the \cielo~galaxies with observations can be found in \citet{Tissera2025}.

Briefly, the \cielo~simulations were run using a version of the P-GADGET-3 code, which includes a multiphase model for the gas component, metal-dependent cooling, star formation and SN feedback, as described in detail in \citet{Mosconi2001}, \citet{Scannapieco2005} and \citet{Scannapieco2006}. Here, we only provide a summary, highlighting the main aspects which are relevant for our work. A more detailed discussion can be found in \citet{Tissera2025}.

The star formation algorithm is implemented in a stochastic fashion depending on the gas density and considering gas clouds with a density, $\rho_{\rm gas} > 10^{-26}{\rm g \ cm^{-3}}$, temperature $T < 15000$ K,  and in a convergent flow, $\vec{\nabla} \cdot \vec{v} < 0$.  A detailed  description is given by  \citet{Scannapieco2005}.
Initially, baryons are in the form of gas with primordial abundance, i.e. X$_{\rm H} = 0.76$, Y$_{\rm He} = 0.24$ and Z $=0$. A \citet{Chabrier2003} Initial Mass Function is assumed in the mass range [$0.1$, $40$] M$_\odot$ \citep{Chabrier2003}. The chemical evolution model includes the enrichment by Type Ia (SNIa) and Type II (SNII) Supernovae\footnote{This set of the \cielo~ simulations do not include AGB enrichment, which are the main producers of C and N. Given that our results are primarily focused on Oxygen abundance, we do not expect this to have a significant impact on our results.}. 

SNIa events are assumed to originate from CO white dwarf (CO WD) binary systems, in which the explosion is triggered when the primary star, due to mass transference from its companion, exceeds the Chandrasekhar limit. For simplicity, the lifetime of the progenitor systems  (delay times) are assumed to be randomly distributed over the range [0.7, 1.1] Gyr. To calculate the number of SNIa,  an observationally motivated relative SNIa rate  is adopted (see also \citet{jimenez2015} for a detailed comparison with the Single Degenerated delay time distribution (DTD) model \citep{Mosconi2001}). The \cielo~project assumes the SNIa yields by \citet{Iwamoto1999}.

SNII are assumed to originate from stars more massive than $\rm 8 M_{\sun}$. The nucleosynthesis products 
 are taken from the metal-dependent yields of \citet{Woosley1995}. The metallicity-dependent lifetimes of SNII are estimated according to \citet{Raiteri1996}.
The code traces the following 12 different chemical elements: H, $^4$He, $^{12}$C, $^{16}$O, $^{24}$Mg, $^{28}$Si, $^{56}$Fe, $^{14}$N, $^{20}$Ne, $^{32}$S, $^{40}$Ca and $^{62}$Zn.

The \cielo~galaxies have been used to study the impact of LG environment onto infall satellites \citep{Rodriguez2022}, the effects of galaxy formation on the shape of dark matter halos \citep{Cataldi2023}, the possible contribution of primordial black holes to the dark matter component \citep{casanueva2024} and the origin of the stellar halos \citep{gonzalezjara2024}.

\subsection{The simulated galaxies}

All \cielo~simulations are analyzed using a consistent methodology as explained in detailed in \citet{Tissera2025}. Briefly, halos are identified at their virial radius, \rvir, using the Friends-of-Friends (FoF) algorithm \citep[][]{davis1985}. Substructures are detected with a {\sc SUBFIND} algorithm \citep{springel2001, dolag2009}, and merger trees are constructed using the {\sc AMIGA} algorithm \citep{amiga}. {We recalculated the center of mass of each substructure by applying the shirking sphere technique.}
For the gas component, we look for cold and dense star-forming regions by selecting only gas particles with temperature $T < 15000$ K and density $\rho > 7 \times 10^{-26}$ g cm$^{-3}$ (hereafter SF particles).  We note that these are the thresholds used by the subgrid physics model to convert gas into star particles together with the requirement for the gas particles to be in a convergent flow.
Hence, we are taking into account all the  cold and dense gas component  because we aim at describing the chemical distribution in the gas that is broadly available for star formation. The larger advantage is that we can better trace the outskirts of discs. Additionally, these results will serve as a baseline to compare with a future work where we mock observations of HII more closely, particulary at high redshift (Tapia-Contreras et al. in preparation).

Galaxies are rotated considering the direction of the total angular momentum of the cold gas. Hence, all metallicity profiles are estimated on the face-on projections. 
We consider as part of a galaxy all mass enclosed within  $2 \times \rm r_{83}$, where $\rm r_{83}$ is the galactocentric radius that enclosed $\sim 83$ percent of the total stellar mass of the system \citep{Tissera2019}. 
We also estimate the stellar half-mass radius, $\rm r_{50}$,  the star formation rate (SFR), considering young stars with ages younger than 0.5 Gyr, and the specific SFR, defined as sSFR$ = {\rm SFR}/$M$_\star$.

\subsection{The oxygen profiles and the DB-A algorithm}
\label{sec:constructionandfit}

\begin{figure}
 \centering
 \includegraphics[width=0.9\columnwidth]{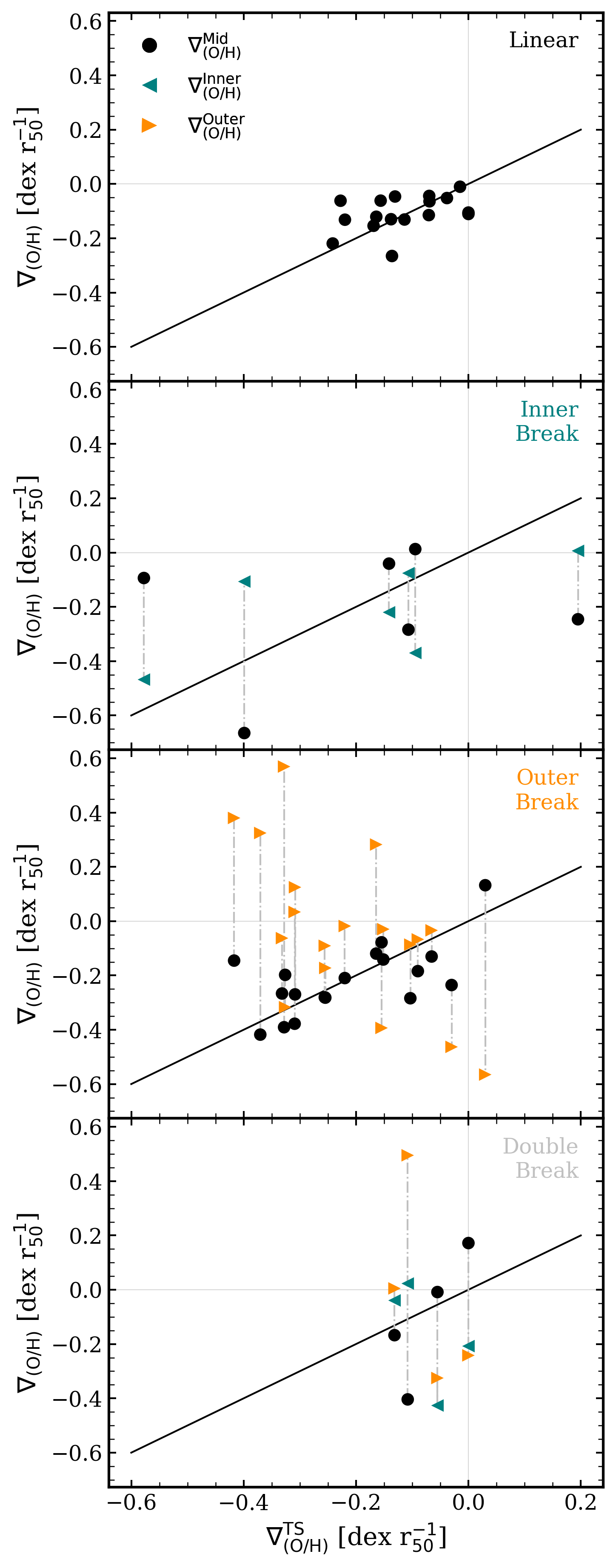}
\caption{
Comparison between the metallicity gradients (filled black circles) estimated by the DB-A fit, \midgrad, and by the Theil-Sen method, \tsgrad, for galaxies exhibiting (from top to bottom): (i) Linear profiles; (ii) Inner Break profiles; (iii) Outer Break profiles; (iv) Double Break profiles. When corresponds, inner (teal symbols) and outer (orange symbols) metallicity gradients are also shown. Vertical lines are included as a visual aid to identify gradients belonging to the same galaxy.
}
 \label{fig:grad_DBA_TS}
\end{figure}

We estimate the metallicity profiles of the gas component as follows.
For each galaxy, we compute the oxygen to hydrogen abundance ratio\footnote{In this work we adopt the usual definition (X/H) where X and H are the number density  of atoms of the corresponding chemical elements.} 12 + log(O/H) for all selected star-forming regions. Galaxies with less than 500 SF particles were discarded. Abundance radial profiles are constructed by computing a running median over radially ordered SF particles, using a window of $N=40$ particles. This significantly reduces the numerical noise associated with the stochasticity of the subgrid methods. However, we are applying azimuthal averages and hence,  azimuthal variations related to the 2D metallicity distribution are not considered \citep{Ho2017, Kreckel2019, Solar2020, Orr2023}.

Because our aim is to study the shape of the profiles and the physical mechanisms behind them, the detection of breaks in the radial profiles needs to be automatically done. This is not a trivial task considering that there is a significant dispersion in the chemical abundances at a given radius as shown in Fig.~\ref{fig:proj+prof}.
A similar situation occurs in observations, as discussed by \citetalias{SanchezM2018} and \citet{Chen2023}, for example. 

For this purpose, we developed an algorithm, hereafter referred to as DB-A, that allows the possible existence of an arbitrary number of breaks. However, based on observational constraints, we set a maximum of two breaks in a given metallicity profile, i.e. an inner break and/or an outer break. No fixed bounds were placed for the break radius, although conflictive cases (i.e. very small or very large break radius or inner and outer break radius too close to each other) were manually revised, decreasing the number of breaks used to fit the profiles. The algorithm applies a Gaussian Kernel Density Estimation\footnote{Which adopts a Scott bandwidth in the form $b = n^{-1/(d+4)}$, where $n$ and $d$ are the number of data and the distribution dimension, respectively.} to estimate the probability density of every particle in the profile and to eliminate outliers if such a value is lower than 0.05. 
DB-A is based on a piece-wise linear function and applies a least-squares method to minimize the error. The algorithm works as follows:
\begin{enumerate}
    \item The maximum number of breaks of the piecewise function $n_{\rm max}$ and a tolerance parameter $\tau$ are defined. We selected $n_{\rm max} = 2$, assuming that the limit case is a doubly-broken linear profile; and $\tau = 0.99$. 
    \item The code iterates over the number of breaks of the piecewise function $n_{\rm br}$, such that $0 \leq n_{\rm br} \leq n_{\rm max}$. The respective piecewise function is fitted in each case, considering scenarios of a single linear trend, a profile broken once, and a profile broken twice. The resultant error\footnote{Root mean square (RMS) error.}, $\varepsilon$, is stored. 
    \item In every iteration we compare the relative errors ($\varepsilon_{\rm n}/\varepsilon_{\rm n-1}$) with our $\tau$ parameter. If $\varepsilon_{\rm n}/\varepsilon_{\rm n-1}$ $<$ $\tau$, the current $n_{\rm br}$ is chosen to be fitted. Otherwise,  the lower value of $n_{\rm br}$ is preferred.
\end{enumerate}

We tested the robustness of our method against both Bayesian
Information Criteria and the Akaike information criterion finding similar results.

We  define inner, intermediate and outer regions according to the detection of breaks. The metallicity gradient, $\nabla_{\text{\scriptsize(O/H)}}$,  is then defined as the slope of the linear function that fits the distribution of chemical abundances in each region,  \innergrad, \midgrad, and \outergrad. We note that when the profile is doubly-broken, both inner and outer gradients  can, therefore, be directly defined. However, when the profile has a single break, $\rm r_{\rm br}$, this can be either an inner or an outer break. Based on the characteristic break radius reported by  \citetalias{Belfiore2017} and \citetalias{SanchezM2018}, our criteria to classify them are the following:

\begin{enumerate}
    \item If $\rm r_{\rm br} < 1.5$ $\times$ $\rm r_{50}$: The break is considered as an inner break, and only inner and middle $\nabla_{\text{\scriptsize(O/H)}}$ are defined,
    \item If $\rm r_{\rm br} > 1.5$ $\times$ $\rm r_{50}$: The break is considered as an outer break, defining only middle and outer $\nabla_{\text{\scriptsize(O/H)}}$.
\end{enumerate}

In the case that a profile is not broken, only the middle gradient is defined, which corresponds to a linear fit to the SF particles within [$\epsilon$, 2 $\times$ \ropt].

In Fig. \ref{fig:proj+prof} we display several examples to illustrate both the diversity of metallicity profiles and the performance of our fitting method. 
In the right panels of Fig. \ref{fig:proj+prof} there are examples of linear, inner break, outer break and doubly-broken profiles. It is important to note that both inner and outer breaks can exhibit either negative or positive gradients. Gas density projections of the simulated galaxies and their environment suggest that the shape of the metallicity profile is affected by the interaction with nearby galaxies as expected. We also show the case of a linear inverted metallicity profile (positive metallicity gradient), which is associated with a clearly disturbed gas distribution \citep[see][]{rupke2010a, fragkoudi2016, sillero2017, Tissera2019}. We will come back to this point later.

For comparison, we also fitted each profile a linear regression (TS)\footnote{We used the Theil-Sen method \citep[TS,][]{theil, sen} to estimate the slope.} within a fixed radial range of $r \in [0.5, 1.5] \times$ \reff. Figure \ref{fig:grad_DBA_TS} shows a comparison between the slope estimated with the two methods, \dbagrad and \tsgrad. We note that for every galaxy we have one gradient estimated with the TS method and up to three gradients estimated with the DB-A method, corresponding to the slope in each region. Figure \ref{fig:grad_DBA_TS} displays \midgrad versus \tsgrad for the four different types of profiles identified. In the case of broken profiles, we also added the corresponding inner or outer slopes. For every type of profile, there tends to be a good agreement between the \tsgrad and \midgrad, favoring a direct comparison between both approaches. Small differences are likely to be related to the adopted radial ranges. 
When the profiles exhibit an inner break, they show the largest discrepancies between the \tsgrad and the \midgrad. 


\section{The gas-phase metallicity gradients in the Local Universe}
\label{sec:redshift0}

\begin{figure*}
\centering
 \includegraphics[width=2\columnwidth]{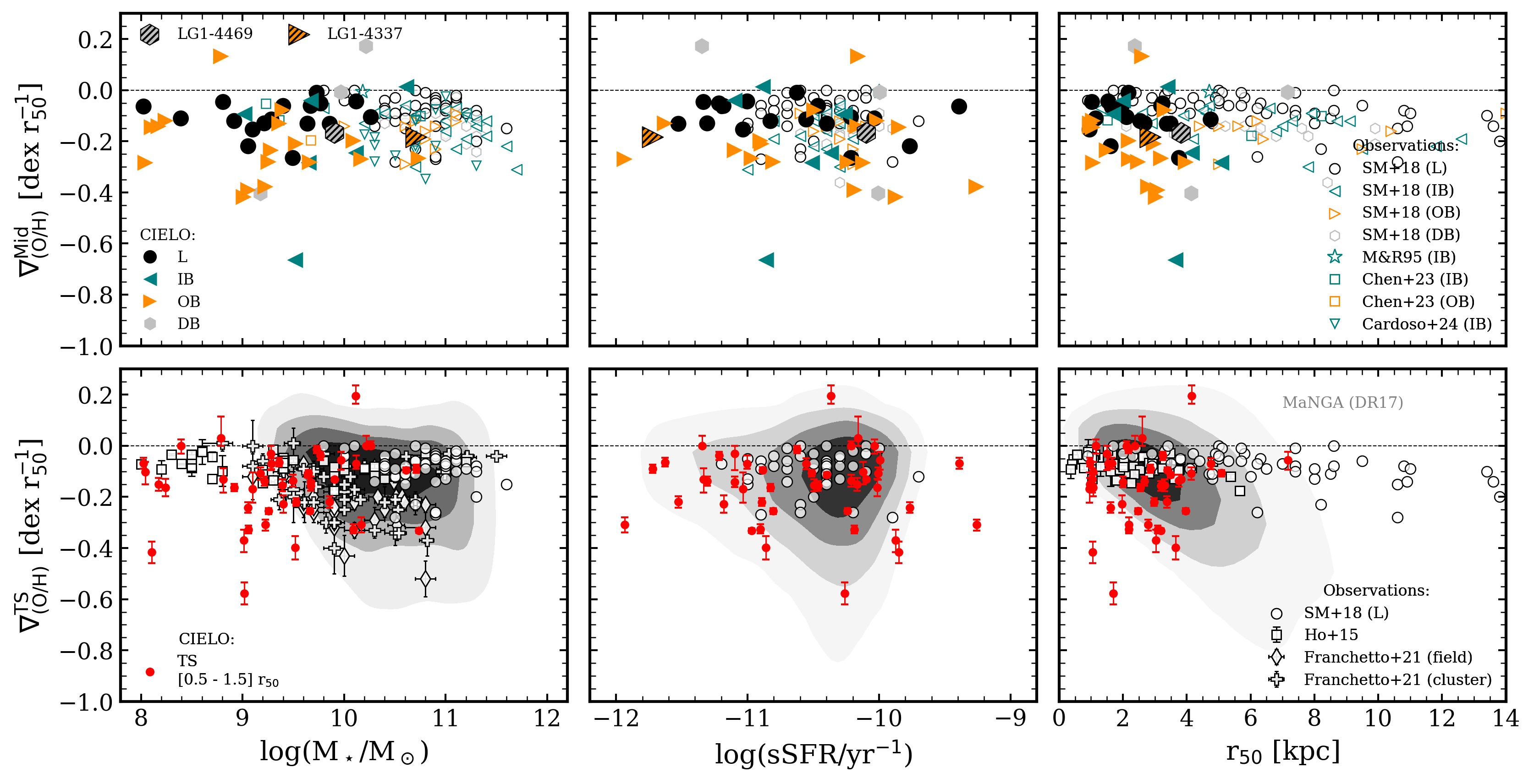}
\caption{Stellar mass-, sSFR-, size-metallicity gradient relation for the 45 star-forming central galaxies (left, middle and right panels,  respectively). Upper panels: the relations are constructed by using only the \midgrad. The colors of solid symbols encode the type of profile that was best fitted to the whole metallicity distribution,  inner break (IB; teal triangles), outer break (OB; orange triangles) and doubly-broken (DB; gray hexagons). Lower panels: The same relations are constructed using \tsgrad (red circles). Observational data from \citet{martinroy1995}, \citet{Ho2015}, \citetalias{SanchezM2018} (MUSE), \citet{Franchetto2021} (MaNGA+MUSE), \citet{Chen2023} (TYPHOON), \citet{cardoso2024} (CALIFA) and MaNGA DR17 are also  displayed in a similar fashion. The black dashed line indicates $\nabla_{\text{\scriptsize(O/H)}} = 0$ for reference purposes only. The two main galaxies LG1-4469 and LG1-4337 of the LG analogue are highlighted as they will be analyzed in detail as case studies.
}
\label{fig:gradmass_cbreaks}
\end{figure*}

\begin{figure*}
\centering
 \includegraphics[width=2\columnwidth]{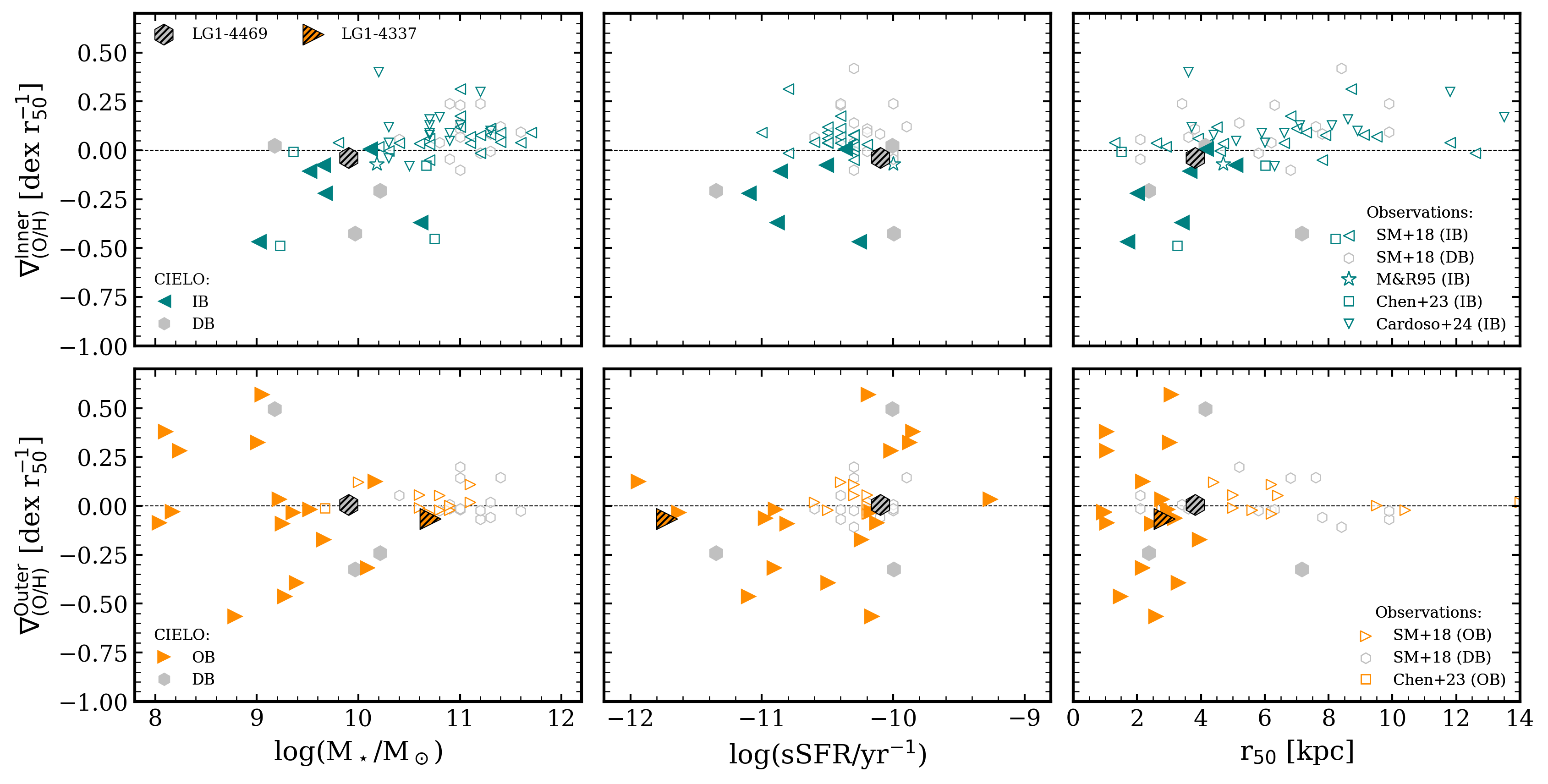}
\caption{Similar to Fig.~\ref{fig:gradmass_cbreaks}, but featuring the \innergrad and \outergrad in broken metallicity profiles. Observational data from \citet{martinroy1995}, \citetalias{SanchezM2018} (MUSE), \citet{Chen2023} (TYPHOON) and \citet{cardoso2024} (CALIFA) are included for comparison.
}
 \label{fig:gradmass_cbreaks2}
\end{figure*}

From the analysis described in the previous section, we found 17 linear (38 per cent), 6 inner-broken (13 per cent), 18 outer-broken (40 per cent), and 4 (9 per cent) doubly-broken profiles at $z=0$. Table \ref{tab:tab_z0} summarizes the best fitting parameters to each profile, including the respective abundance gradients in each region and the break radii if corresponds. 
In the left, upper panel of Fig.~\ref{fig:gradmass_cbreaks} the MGMR is displayed using the metallicity gradients of the intermediate regions, \midgrad.  The different symbols and color code indicate the global shape of the profiles (i.e. inner, outer or no breaks). 
For comparison the metallicity gradients fitted by \citetalias{SanchezM2018} for MUSE galaxies \textcolor{black}{and by \citet{cardoso2024} for CALIFA galaxies} are also displayed in a similar fashion. 
Additionally, the metallicity gradients reported by \citet{Chen2023}, obtained by fitting a broken profile to bar disc galaxies, are also included.  
The \midgrad(\dexre) of the \cielo~galaxies generally aligns with the observations. However, a larger variety is detected  for simulated galaxies with an outer broken profile, a feature observed across the entire range of stellar masses. \citetalias{SanchezM2018} found 53 per cent of their sample to show linear negative gradients while  21 per cent have inner breaks, 10 per cent only outer breaks and 17 per cent are doubly broken. These frequencies differ from those obtained for the \cielo~galaxies, where the majority exhibit breaks.  We estimated the binomial probability of our results given the \citetalias{SanchezM2018} rates, and found a low probability (p<0.01), which indicates that the frequency of breaks in CIELO is larger than expected from \citetalias{SanchezM2018} observations. 
We note that the stellar mass range covered by the \cielo~galaxies is different from that of SM18 and this could be an important factor \citep{Belfiore2017}. Additionally, this difference could also reflect the fact that in the simulated galaxies we used the whole cold and dense gas distribution  while observations are restricted to detections limits, which can prevent to measure the outer breaks, for example. Extending both the observed and the simulated samples to cover similar mass range of galaxies could allow a more robust assessment of the differences, which could also provide hints on galaxy formation models and the adopted subgrid physics.

In the lower, left panel of Fig.~\ref{fig:gradmass_cbreaks}, the MGMR is displayed using \tsgrad for the \cielo~sample. For comparison, observations reported by  \citetalias{SanchezM2018}, i.e. those which follow a linear profile, and by  \citet{Franchetto2021}, \citet{Ho2015} and \citet{Belfiore2017}, where a linear regression is fitted.
As can be seen, the simulated values are within the observed range. There are some galaxies which show steeper negative gradients but, nevertheless, within the observational range. 
In the next section we further discuss the origin of the inverted/positive slopes in the metallicity distributions.

In the middle and right panels of  Fig.~\ref{fig:gradmass_cbreaks}, we display the \midgrad (upper panels) and \tsgrad (lower panels)  as a function of sSFR  and r$_{50}$, respectively.
As can be seen the relation between \midgrad and both parameters are within the observed range. We note that \cielo~galaxies are smaller compared to the \citetalias{SanchezM2018} as expected because of the  differences in stellar mass as can be seen from the left panels of this figure.
At $z=0$ we find only three galaxies with positive \midgrad (see Table \ref{tab:tab_z0}). These galaxies show signals of recent interactions.

Regarding the inner and outer slopes, we performed similar relations and include available data to confront them. In Fig.~\ref{fig:gradmass_cbreaks2}, the corresponding relations for the \innergrad (upper panels) and \outergrad (lower panels) are displayed. As can be seen the simulated values are within the observed range. In the case of the inner drops, they match the correlation signal reported by \citet{cardoso2024}. For low stellar mass galaxies, inner rises or negative slopes appear to be preferred. The simulated slopes are consistent with observations reported in \citet{Chen2023}. 

The lower panels of Fig.~\ref{fig:gradmass_cbreaks2}
displays the same relation but for the \outergrad. There is a significant dispersion that decrease with stellar mass. Observations and simulations report similar slopes for galaxies with M$_\star > 10^{9.5}$ \msun. Simulated galaxies with positive \outergrad tend to have more active star formation. Positive slopes in the outer region could be linked to accretion of pre-processed material or satellite interactions \citep{perez2011}. We will discuss these different physical processes by using case studies in the following section.

\begin{figure*}
\centering
 \includegraphics[width=2\columnwidth]{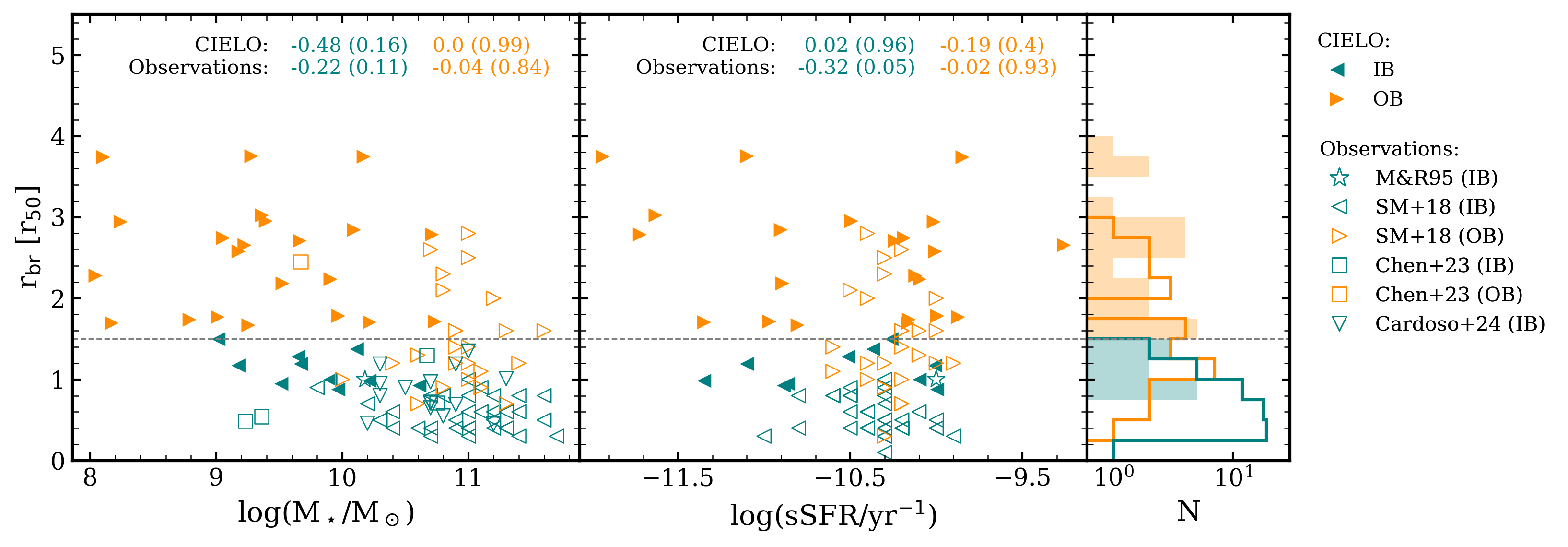}
 \caption{Inner (teal symbols) and outer (orange symbols) break radii as a function of: (i) galaxy stellar mass (left panel); (ii) sSFR (right panel). \cielo~galaxies are displayed as filled symbols, while empty symbols represent observations from \citet{martinroy1995}, \citetalias{SanchezM2018}, \citet{Chen2023} and \citet{cardoso2024} as indicated in the legend. The Spearman Correlation Coefficient (SCC) was calculated to evaluate possible correlations between galaxy properties and inner and outer break radii. The upper part of each panel shows the SCC computed considering: \cielo~galaxies (top rows) and observations (bottom rows). The p-value is shown in parenthesis in each case.}
 \label{fig:rbr_z0}
\end{figure*}

Finally, in Fig.~\ref{fig:rbr_z0}, we display the characteristic inner and outer radii as a function of stellar mass and sSFR. We also included observations from \citetalias{SanchezM2018}. The simulated inner breaks have a median (25$^{\rm th}$ -- 75$^{\rm th}$ percentiles) of 1.08 (0.96 -- 1.26) \reff and outer breaks of 2.62 (1.77 -- 2.912) \reff. 
We do not detect any statistically significant trend with stellar mass or sSFR for the simulated or the observational data as quantified by the Spearman Correlation Coefficient (SCC), except for the observed inner breaks, which seems to define an anti-correlation with sSFR \citep[see also][]{cardoso2024}.

In  Fig.~\ref{fig:gradmass_cbreaks} and Fig.~\ref{fig:gradmass_cbreaks2} we have highlighted with larger symbols two galaxies,  LG1-4469 and LG1-4337. These target galaxies will be used as examples to analyze the impact of different physical mechanisms on the shape of the metallicity profiles.
At $z=0 $, LG1-4469 and LG1-4337 have a double break and an outer break, respectively.

\section{Case studies on the evolution of gas-phase metallicity gradients}
\label{sec:evolution}

In this section, we trace LG1-4469 and LG1-4337 from the  LG1 analogue, back in time along their merger trees, using them as case studies.  We analyze  the details of their evolution to gain insights into the physical processes that can generate these breaks and the timescales over which these profiles change shape. We expect the evolution of the metallicity gradients to  reflect their distinct evolutionary histories \citep[e.g.][]{ma2017,Tissera2021}, but to share common features.
At each available snapshot, the same procedure applied to galaxies at $z=0$ (see Sec. \ref{sec:constructionandfit}) is performed on their progenitors up to $z \sim 6$. Hence,  between $z=0$ and $z\sim 6$ this analysis was applied to 87 available snapshots\footnote{We acknowledge the use of the \cielo~database \citep{gonzalezjara2024}.}. We remind the reader that only systems with more than 500 SF particles are considered.

At all available redshifts, we rotate the gaseous discs along the direction of total angular momentum, estimate the projected metallicity gradients and applied the DB-A  method to all of them. Hence, at each $z$ we have the \innergrad, \midgrad and \outergrad with the corresponding breaking radii. Additionally, we estimated   \ropt and \reff for the progenitors of both galaxies.

We  define the regions associated with breaks or equivalent ones  to evaluate the impact of the acting physical processes. For this purpose, we define the inner, middle and outer regions as follows:

\begin{enumerate}
    \item Inner region: $\rm r \in [0, r_{\rm inner})$ kpc
    \item Middle region: $\rm r \in [r_{\rm inner}, r_{\rm outer})$ kpc
    \item Outer region: $\rm r \in [r_{\rm outer}, $ $2\times \rm r_{83}]$ kpc
\end{enumerate}

\noindent
where 
\begin{equation}
{\rm r}_{\rm inner} = 
\left\{
    \begin{array}{lr}
        {\rm r}_{\rm br}, & \text{if the profile has an inner break at }{\rm r}_{\rm br} \\
        {\rm r}_{50}, & \text{otherwise}
    \end{array}
\right.
\end{equation}

\noindent ,
\begin{equation}
{\rm r}_{\rm outer} = 
\left\{
    \begin{array}{lr}
        {\rm r}_{\rm br}, & \text{if the profile has an outer break at }{\rm r}_{\rm br} \\
        {\rm r}_{83}, & \text{otherwise}
    \end{array}
\right.
\end{equation}

In each of these regions, we  computed the SFR  in a similar way as explained in the previous section. Since inner, middle and outer regions can differ in size during the evolution of the galaxies, we also calculate the SFR surface density, $\Sigma_{\rm SFR}$, within each of the three defined regions (see Appendix~\ref{appendix1}).

\subsection{Environmental and secular processes}
\label{subsec:mergersandsatellites}

To understand the physical processes modulating the shape of the metallicity profiles, we explored the relationship between the central galaxy and its environment by following the accretion of gas from different sources and the interaction and merger histories. Satellite interactions and merger events are mechanisms that affect the galaxy dynamics and distribution of both gas and stars, in addition to being able to contribute gas for star formation as discussed in the Introduction. We also search for gas particles accreted from the circumgalactic medium (CGM) \citep{Peroux20} and from cold flows \citep{ceverino2016}.

\subsubsection{Surviving satellites}
\label{subsec:satellites}
In order to study the  effects on the abundance distributions associated to surviving satellites, we identified all the substructures surrounding the central galaxy at $z=0$, and select as satellite galaxies those whose center of mass are within \rvir of the central galaxy, and have a stellar mass $\log({\rm M_\star/M_\odot}) > 7$. An interaction is defined from the time a satellite enters the virial radius of a central galaxy. We followed their evolution back time and obtained their stellar masses, gas fractions and distance to the center of mass (d$_{\rm CM}$) of the progenitors of our two target  galaxies up to $z\sim 6$. These properties are displayed in detail in Appendix \ref{app:mergerhistories} as a function of time. The most significant periods of interaction, in terms of stellar mass fraction and gas content, are highlighted  in the upper panels of Fig.~\ref{appfig:mergers4469} (dashed, green regions).

\subsubsection{Mergers}
\label{subsec:mergers}
Furthermore, we followed the merger tree of each central galaxy to identify the most significant merger events that occur during each galaxy assembly process. We selected mergers with a stellar mass ratio ${\rm M}_{\star, \rm merger}/{\rm M}_{\star, \rm central} > 0.01$ at the infall time (i.e. when the center of mass of the merging galaxy crosses \rvir of the central galaxy progenitor). We define a merger event when the infalling satellite is not longer identified as a separate substructure by the SUBFIND algorithm. In terms of the gas content of the merging galaxies, we considered both wet and dry mergers. A high amount of gas can be accreted into the central galaxy during wet mergers, while the torque exerted by dry mergers can also have an impact on the redistribution of gas within the central galaxy. The properties of such mergers are displayed in detail in Appendix \ref{app:mergerhistories}. Analogous to satellite interactions, the selected merger events are highlighted in the lower panels of Fig. \ref{appfig:mergers4469} (dashed, blue regions).

\subsubsection{Accretion rates}
\label{subsec:accrates}

\begin{figure*}
 \centering
 \includegraphics[width=2\columnwidth]{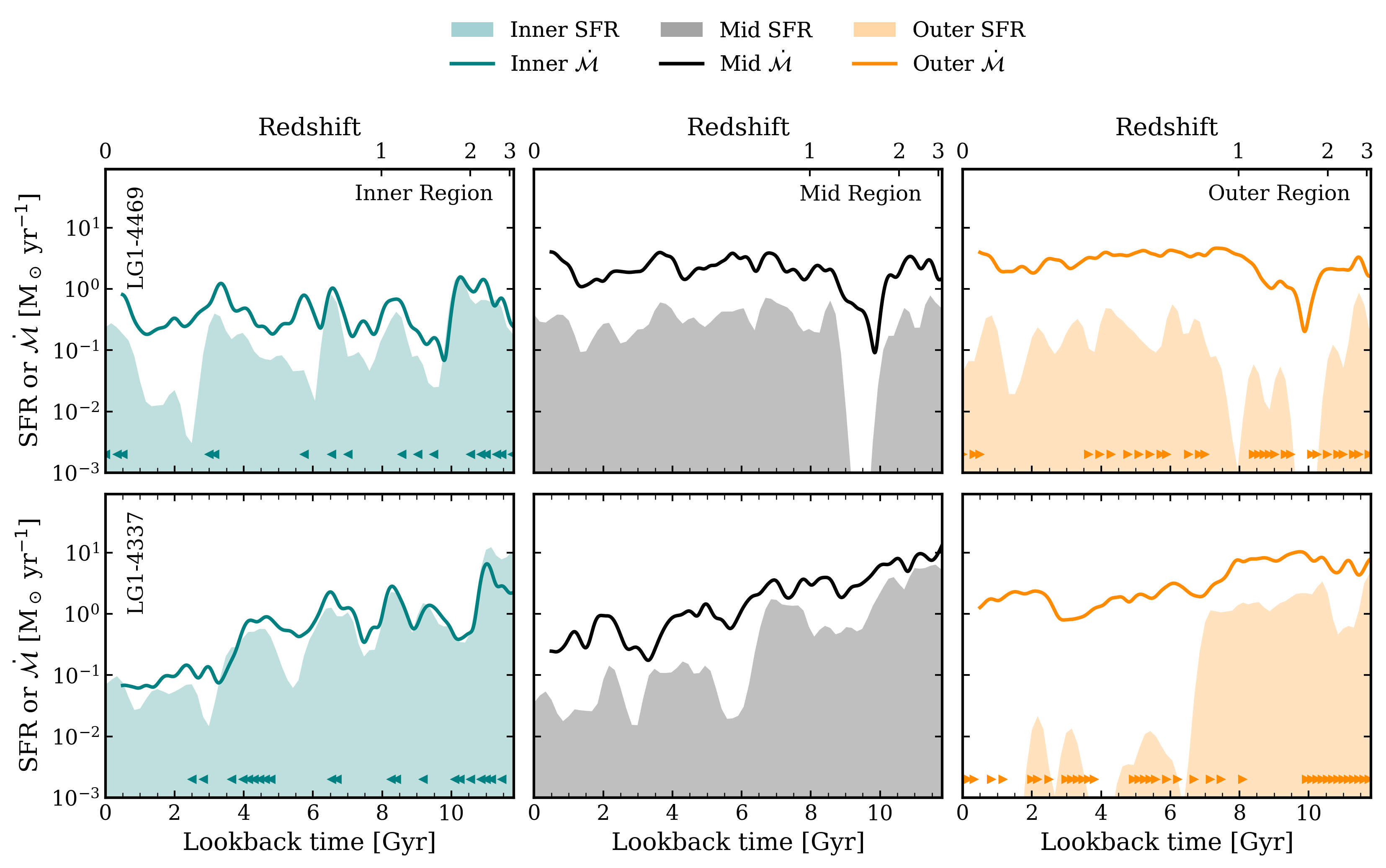}
 \caption{Time evolution of the SFR (shaded areas) and the total accretion rates ($\Dot{\mathcal{M}}$; solid lines) per region (inner: left column; mid: central column; outer: right column) for each galaxy (LG1-4469: top row; LG1-4337; bottom row). The epochs where inner or outer  breaks are identified are denoted by using teal or orange symbols, respectively (see Section \ref{sec:evolution} for a detail explanation on the definition of the different regions).
Figure \ref{fig:ARsintime_norm} shows the same quantities normalised by the area of each region.
 }
 \label{fig:ARsintime}
\end{figure*}

\begin{figure*}
 \centering
 \includegraphics[width=2\columnwidth]{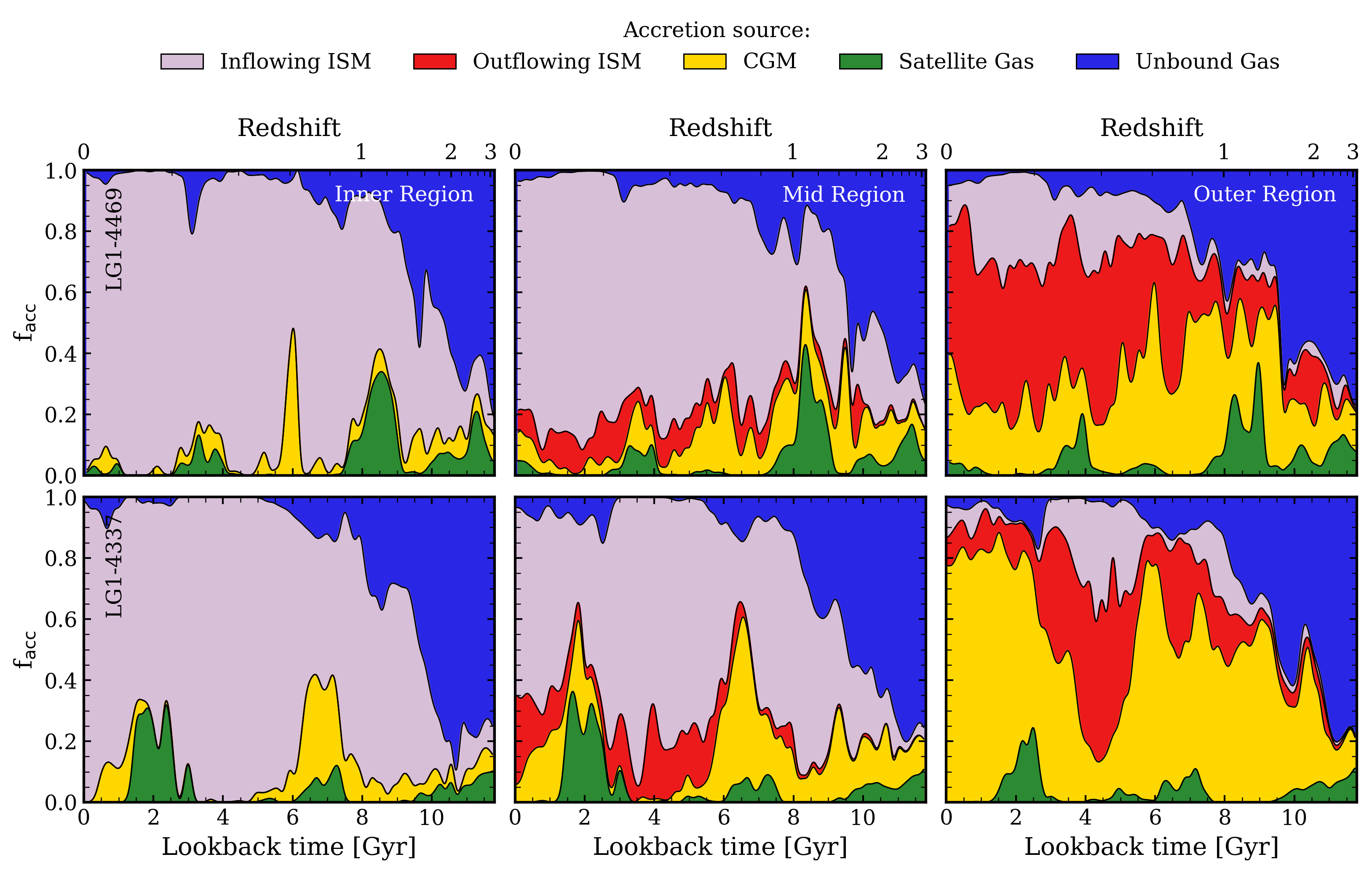}
 \caption{Fractions of the total accreted gas ($\Dot{\mathcal{M}}$; see Fig. \ref{fig:ARsintime}) in the  inner (left column), middle (central column) and  outer (right column) regions for each galaxy (LG1-4469: top row; LG1-4337; bottom row). The gas accreted fractions correspond to inflowing ISM (pink areas), outflowing ISM (red areas), the CGM (yellow areas), interacting/merging satellite (green areas) or unbound gas (blue areas).}
 \label{fig:AR_fractions}
\end{figure*}

In order to understand the gas motion and infall into the target galaxies, we estimated three gas accretion rates, $\Dot{\mathcal{M}}$, one for each region defined at the beginning of Sec. \ref{sec:evolution}. The accretion rates are calculated as the gaseous mass that enters the inner, middle or outer regions per unit Gyr (rates per unit area are provided in Section~\ref{appendix1}). We also considered five different accretion sources. To do this, at a given time, the gas particles that currently belong to the central galaxy, but did not belong to its progenitor in the previous available snapshot, are followed back in time for $\sim 1$ Gyr to check if: they belonged to another galaxy (hereafter, satellite gas); they were not gravitationally bound to any structure (hereafter, unbound gas); or they were part of the gas surrounding the progenitor or CGM, defined as the  gas located further than $2 \times r_{83}$) but within the \rvir. 

In addition, we also estimated the redistribution of gas within a galaxy (ISM) as the gaseous mass that moves from one region to another, in any direction. Radial and vertical displacements of gas particles between two consecutive snapshots would be noted as $\delta r$ and $\delta z$, respectively. We discarded radial displacements smaller than the gravitational softening ($\epsilon_{\rm g}\sim 0.4$ physical kpc for LG1 at $z=0$). 
As a convention, we will use $\delta r < 0$ to refer to particles moving outwards (hereafter, gas radial outflows), and in the opposite case, $\delta r > 0$ to refer to particles moving inwards (hereafter, gas radial inflows). 

\subsection{Evolution of the star formation rate and gas accretion}

Figure~\ref{fig:ARsintime} shows the temporal evolution of both the SFR and \accrates in units of \msunperyr.Fig.~\ref{fig:ARsintime_norm} shows the evolution of the $\Sigma_{\rm SFR}$ and $\Sigma_{\Dot{\mathcal{M}}}$ per region (i.e. SFR and \accrates normalized by the area of each region). As can be seen, both quantities are strongly correlated in the inner region, particularly when there is active star formation. This suggests that the accreted gas is efficiently condensed and converted into stars. In the middle and outer regions, both quantities evolve similarly but the amount of accreted gas tend to be $\sim 1-2$ orders of magnitude higher than the SFR. This behavior can be explained by the lower gas density with respect to the central regions. This is consistent with the low  star formation efficiency typically observed in the outskirts of galaxies compared to their central regions \citep{leroy2008, bigiel2008, krumholz2009}. It suggests that gas accumulates in the disc and is steadily converted into stars over time. At low redshift, the outer region of LG1-4337 provides an example of active accretion with very low star formation activity, emphasizing the potential influence of additional physical processes beyond accretion in regulating star formation activity in galactic outskirts, thereby adding to the complexity (e.g. turbulence).

In order to understand the origin of the gas accretion, in Fig.~\ref{fig:AR_fractions} we display contribution of each of the defined accretion sources. In the inner region, the dominant accretion source transitions from unbound gas (blue areas) to inflowing ISM (pink area), for decreasing redshift. this indicates that at later times, most of the gas that reaches the central regions is primarily accreted from the ISM reservoir in the discs. Peaks in accretion from satellites and the CGM can results in up to f$_{\rm acc} \sim 0.4-0.5$, often driven by interactions that either directly transfer gas from companions or trigger gas inflows through angular momentum loss from the  disc ISM (pink areas) and CGM (yellow areas).

The middle regions show a more diverse origin for accreted gas, although the infall of ISM from the outermost disc remains the dominant source (pink areas), typically contributing between 0.3 and 0.8 to the fraction of gas accretion. Surges of accreted gas from satellites and the CGM, comparable to those observed in the inner region, occur within the same time intervals. Additionally, outflowing material of the ISM (red areas), potentially redistributed by SN feedback or the disturbance of the discs produced by tidal torques during close encounters, contributes up to f$_{\rm acc} \sim 0.3$ \citep[see also][]{sillero2017}.

The outer regions show the most diverse sources of accretion, with distinct patterns observed in the two case studies. The high-mass galaxy, LG1-4337, displays a smooth transition where unbound gas is gradually replaced by CGM gas as the dominant accretion source. This CGM gas has been likely processed in  central region of the galaxy, later ejected by SN feedback to feed the CGM, and subsequently re-accreted due by the deep gravitational potential of the galaxy \citep{Tumlinson17, Peroux20, peroux2020}. Peaks in accretion of radially outflowing gas correlate with mergers (see Fig. \ref{appfig:mergers4469}) that can disrupt part of the external spiral arms and trigger star formation bursts (see Fig. \ref{fig:ARsintime}) that subsequently drives galactic ouflows (Miranda et al. in preparation).

For the low-mass galaxy, LG1-4469, the transition begins with unbound gas being replaced by CGM gas (lookback time of 10–7 Gyr) and later by outflowing ISM gas (lookback time: of  7–5 Gyr). During the first period, the galaxy is disrupted by a merger event (see Fig.~\ref{appfig:mergers4469}), with CGM accretion resulting from the relaxation of the galaxy and rapid infall of surrounding gas remnants. 

In the second period, the contribution of the ouflowing ISM becomes significant. This corresponds to radially heated particles and the opening of spiral arms caused by interactions with a surviving satellite (see Fig.~\ref{appfig:mergers4469}).
Both galaxies also receive significant contributions from the accretion of satellite gas during close encounters, as previously observed in the inner and mid regions.

Globally, at earlier times, unbound gas, representing cold flows from cosmic filaments, contributes significantly across all regions. This source dominates (f$_{\rm acc} \geq 0.7$ ) until approximately $z \sim 1$ in the outer regions and $z \sim 1.5 $ in the other two inner regions. Later on,  f$_{\rm acc}$ decreases down to less than 0.05.

While each galaxy exhibits a unique accretion pattern in each region, reflecting the interplay between their internal properties and the characteristics of their environment over time, the physical processes contributing gas remain the same. Hence what varies with redshift is their relative importance and the properties of the accreted gas. In the following section, we explore the impact of these physical processes on the shape of the metallicity gradients.

\subsection{Evolution of the metallicity gradients}

\begin{figure}
\centering
 \includegraphics[width=\columnwidth]{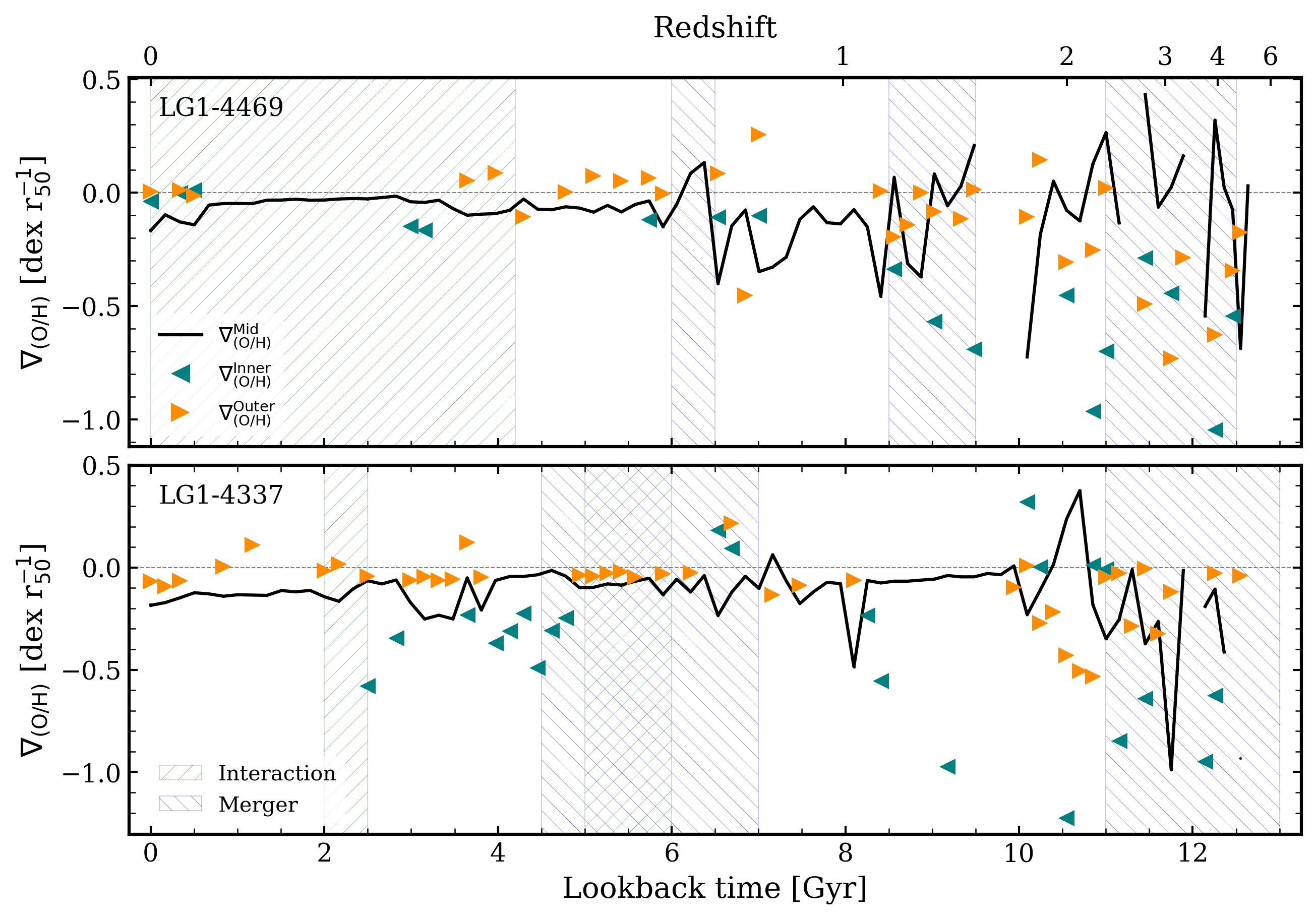}
 \caption{Evolution of the metallicity gradients as a function of lookback time (and redshift) for the targeted galaxies LG1-4469 (upper panel) and LG1-4337 (lower panel). The metallicity gradients in the inner (teal symbols), middle (black solid lines) and outer (orange symbols) regions are displayed. The shaded regions depict the mergers and interaction events (blue and green, respectively).}
 \label{fig:gradientsintime}
\end{figure}

Figure~\ref{fig:gradientsintime} illustrates the evolution of the gradients for both targeted galaxies across time. We show, when corresponds, the gradients in the inner (teal symbols) and outer (orange symbols) regions defined by the DB-A method, as explained in Section~\ref{sec:constructionandfit}. The gradients are normalised by the corresponding \reff at every time. The general and consistent behavior observed in these galaxies is that the metallicity gradient undergoes substantial variability at high redshift, progressively transitioning to a more gradual evolution at lower redshifts, with this trend being particularly evident in the middle gradient. 
This behavior is consistent with previous works \citep[e.g.][]{ma2017, Tissera2021}, indicating  higher variability of the metallicity gradients associated to higher frequency of merger and gas-richness of galaxies at higher redshift.

In Fig.~\ref{fig:gradientsintime}, the shaded regions indicate periods when the target galaxy is involved in an interaction with  surviving satellites (green regions) or undergoing  a merger event (blue regions). These events were identified and selected as described in Sec. \ref{subsec:satellites} and \ref{subsec:mergers}, respectively. The full merger history is shown and discussed in the Appendix (see Fig. \ref{appfig:mergers4469}), including the information about the gas content of the interacting galaxies and its mass ratio with respect to the central galaxy. It can be seen that the periods of greater variability in the metallicity gradient are linked to these events, with a stronger impact at high redshift due to the higher gas fraction and relative mass ratio of the galaxies involved. 

Inner and outer gradients are also steeper at higher redshift and tend to be more negative. It is clear from  Fig.~\ref{fig:gradientsintime} that, at least for the target galaxies, \outergrad tends to be negative at higher redshift but as the galaxies evolved they became flatter or even positive. In the case of \innergrad, most of them are negative and only in few occasions, they became positive. 

In the following sections, we will focus the discussion on the physical processes behind the occurrence of inner and/or outer breaks. As shown in Fig. \ref{fig:grad_DBA_TS} and previously discussed in Sec. \ref{sec:redshift0}, the \midgrad is similar to the gradient estimated with more conservative methods assuming a single linear function. Hence, we  further investigate the global evolution of the gradients in this region, which has been extensively discussed in previous works \citep{Gibson2013, ma2017, yates2020, Tissera2021, hemler2021, vallini2024, venturi2024}. Instead,  we focus on the processes that could originate outer and inner breaks.

As shown in Fig. \ref{fig:gradientsintime}, inner breaks occur at different epochs throughout the entire analyzed redshift range. The two target galaxies exhibit a total of 46 inner breaks (21 for LG1-4469 and 25 for LG1-4337), representing 26 per cent of the total analyzed metallicity profiles. These breaks can appear either as single break or as part of doubly-broken profiles. 

\begin{enumerate}
    \item Inner Rise: The inner gradient is negative and steeper than the mid-region gradient $\bigl($\innergrad < \midgrad$\bigr)$.
    \item Inner Drop: The inner gradient is flatter or inverted compared to the mid-region gradient $\bigl($\innergrad > \midgrad$\bigr)$.
\end{enumerate}

Using this classification, we found that 72 per cent of inner breaks correspond to inner rise profiles, which are characterized by high central metallicity and a steep negative slope in the inner region. The remaining 28 per cent are inner drop profiles, which display diluted central metallicities, resulting in a flattened or inverted metallicity gradient in the inner region.

We find outer breaks to be the more common feature across time, being present in a total of 75 metallicity profiles (35 for LG1-4469 and 40 for LG1-4337), 43 per cent of the complete sample. 

We find two scenarios for the outer region: 
\begin{enumerate}
    \item Outer Rise: The outer gradient is flatter or inverted compared to the mid-region gradient $\bigl($\outergrad > \midgrad$\bigr)$.
    \item Outer Drop: The outer gradient is negative and steeper than the mid-region gradient $\bigl($\outergrad < \midgrad$\bigr)$.
\end{enumerate}

Outer rises, characterized by high metallicities in the outskirts, are found in 68 per cent of the cases. Outer drops are present in the remaining 32 per cent, and they are found almost exclusively at high redshift (88 per cent at $z>1$; see Fig. \ref{fig:gradientsintime}).


\section{Discussion on the origin of inner and outer breaks}
\label{sec:discussion}

In this section we discuss the origin of the inner and outer breaks using the analysis we have performed so far. We are applying this to our two case studies to illustrate the action of the mentioned physical processes. There might be variations on the response of the metallicity distribution to these effects depending on the specific conditions of the progenitors and their environment. This detailed analysis will be considered as a baseline for an statistical study, which be conducted in a separate paper, and will consider galaxies in different environments.

\subsection{Inner rises}
\label{subsec:innerrises}

\begin{figure*}
 \includegraphics[width=2\columnwidth]{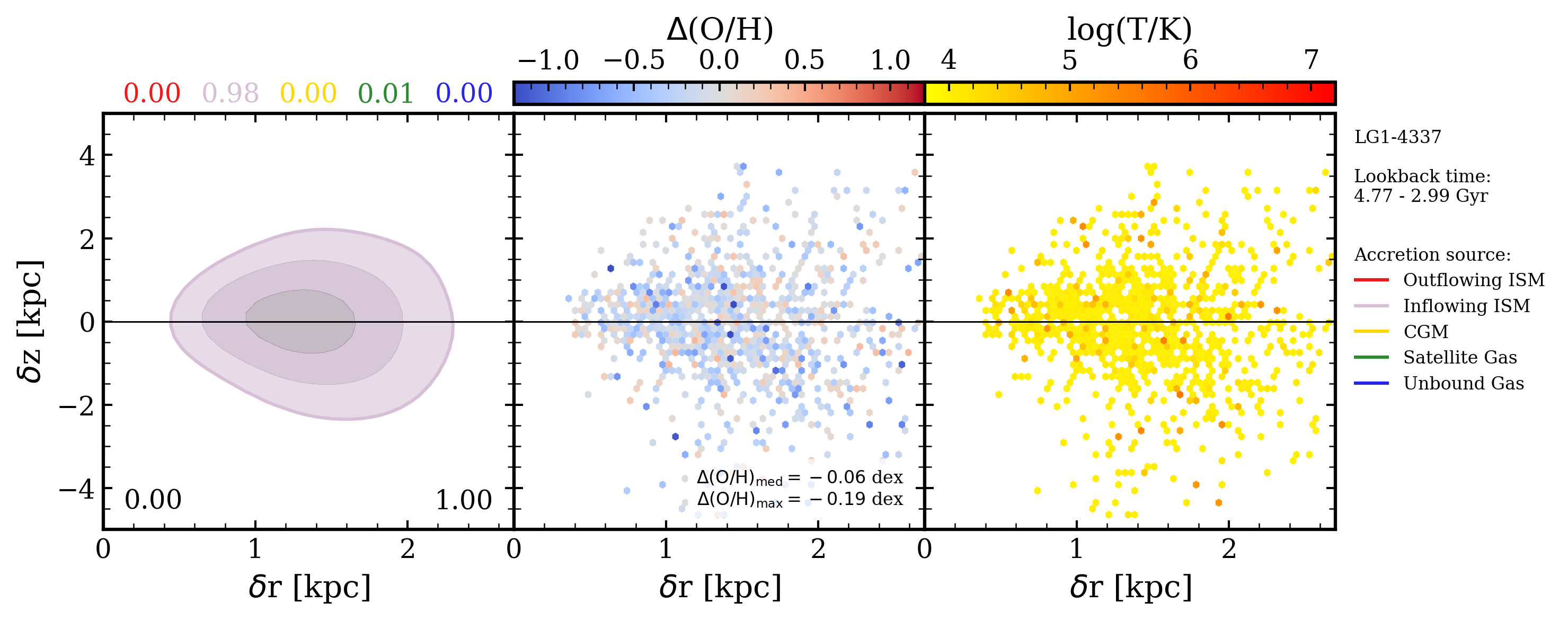}
 \caption{$\delta r - \delta z$ diagram illustrating the motion of the gas accreted into the inner region of LG1-4337 between lookback time $\sim 4.8$ Gyr and $\sim 3$ Gyr. During this period, the galaxy exhibits an inner break of its metallicity profile. Left panel: KDE distribution of the accreted particles. The colored numbers displayed at the top represent the fractional contribution of each gas accretion source, consistent with the color code used in Fig. 7 and the legend. The numbers below indicate the fraction of particles with $\delta r < 0$ (outflowing) and $\delta r$ > 0 (inflowing). Central panel: distribution of accreted particles, color-coded by $\Delta{\rm (O/H)}$, which represents the difference between the metallicity of the accreted gas and that of the respective region. Negative values indicate that the accreted gas is less enriched than the pre-existing gas in the region. The median (med) value is computed over the entire time interval, while the maximum (max) value corresponds to the largest difference observed within that period. Right panel: distribution of accreted gas particles colored by their temperature.}
 \label{fig:drdz1}
\end{figure*}

\begin{figure*}
 \includegraphics[width=2\columnwidth]{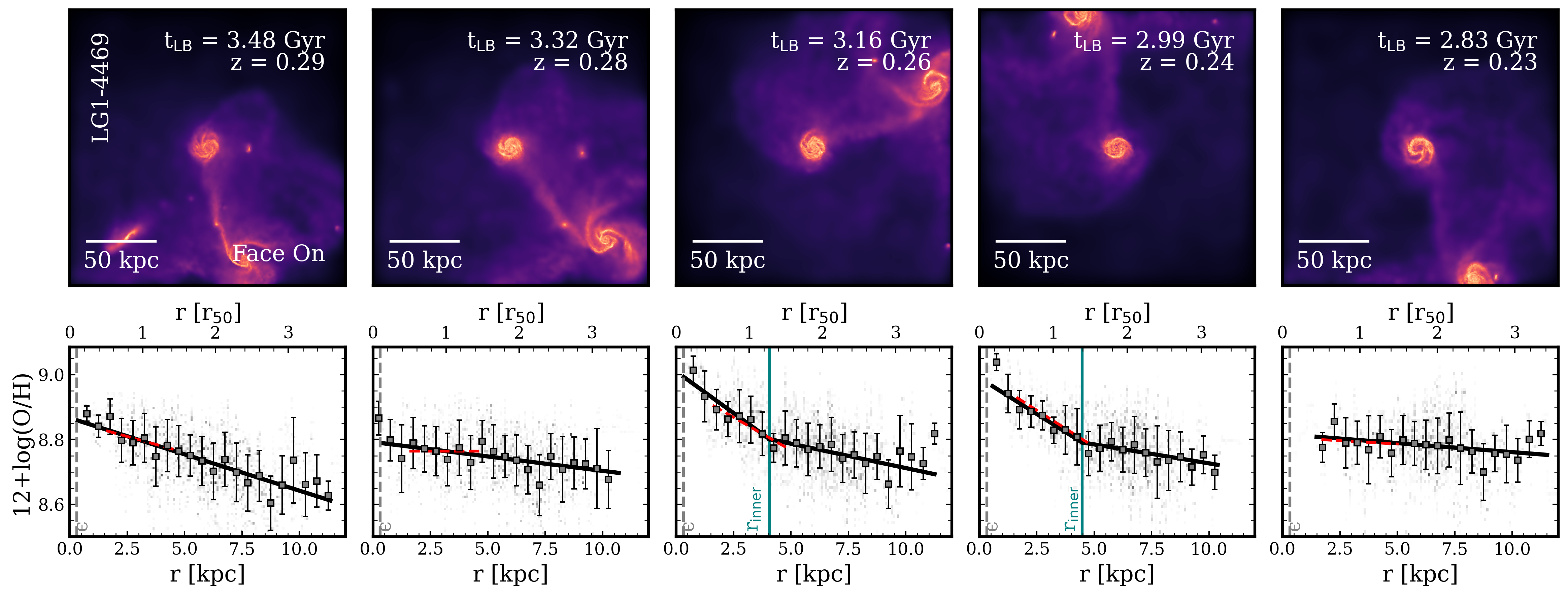}
 \includegraphics[width=2\columnwidth]{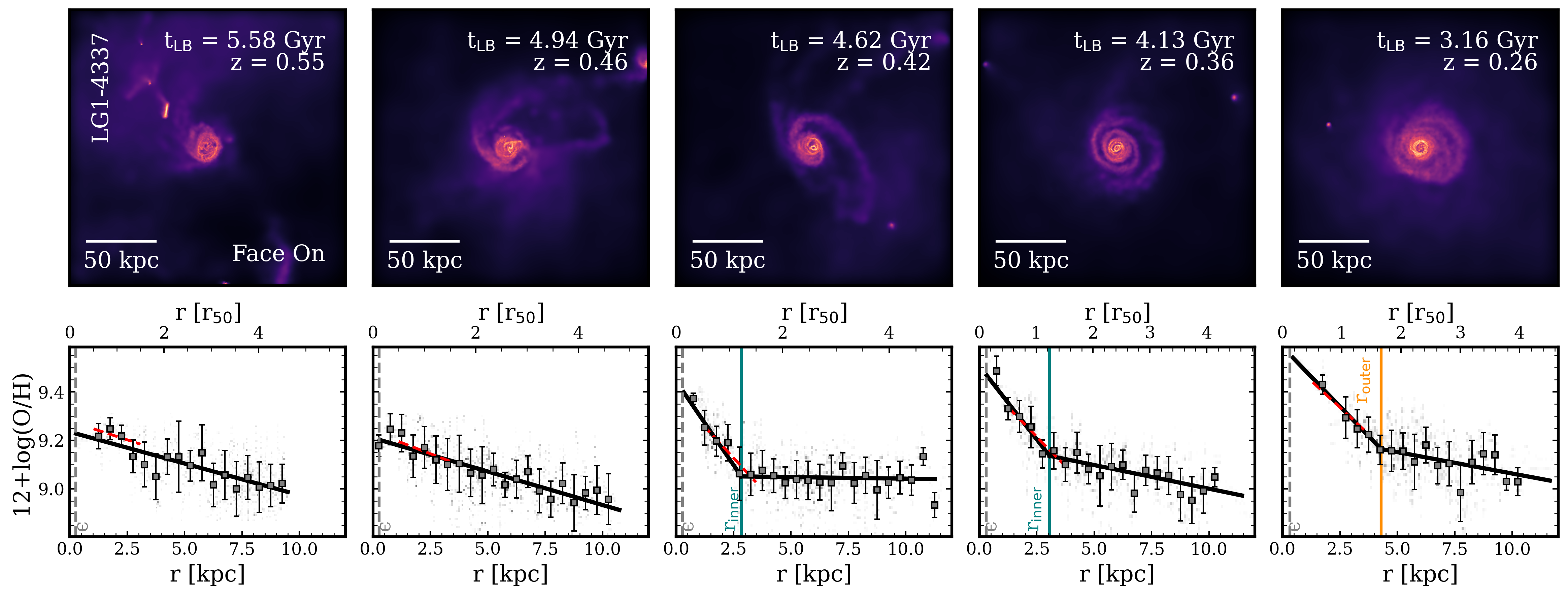}
 \caption{Temporal sequence showing the evolution of inner rise metallicity profiles for LG1-4469 (first and second rows) and LG1-4337 (third and fourth rows). The labeled images depict gas density projections at different snapshots, while the corresponding metallicity profiles are displayed below, following the same format as in Figure \ref{fig:proj+prof}. Note that the timescales differ between the two galaxies.}
 \label{fig:innerrise}
\end{figure*}

From Fig.~\ref{fig:AR_fractions} we can see that the inner region receives gas from two main sources, the ISM at low redshift and unbound gas at high redshift. Additionally, from Fig. \ref{fig:ARsintime_norm}, we can appreciate that the inner region displays the highest accretion rates per unit area compared to the middle and outer regions. 

Most of this material originates from ISM gas moving into the  center of the galaxy, driven by tidal torques produced in galaxy interactions (see Fig.~\ref{appfig:mergers4469}).
While the unbound gas is found to fall along filamentary structures. 

In Fig.~\ref{fig:drdz1} we display the  motions, the relative level of enrichment and temperature of the accreted gas according to their origin during a key period of evolution  when the profiles exhibit an inner rise. The gas motion is quantified by $\delta r$ and $\delta z$ (see definition in Section \ref{subsec:accrates}). The relative level of enrichment, $\Delta{\rm (O/H)}$, is  defined as the difference between the metallicity of the accreted gas and that of the respective region. Negative values indicate that the accreted gas is less enriched than the pre-existing gas in the region. Finally, the temperature distribution of the accreted material.

Overall, the dominant process that change the inner slopes is the infalling of cold gas from the disc ISM, with lower metallicity than the central regions \citep{kewley2006, ellison2013, bustamante2020}. We find that this gas inflows first dilute the central metallicity but soon after they also trigger star formation which injects new elements producing a negative increase of the metallicity gradient as shown by previous works \citep{perez2011,Torrey2012}. 
In fact, \citet{sillero2017} observed a similar pattern in simulations of interacting galaxies, where interactions induced gas inflows, boosting SFR and diluting central metallicity, leading to flatter gradients.

The impact of mergers is detected at high and low redshift with the difference that there is larger frequency of wet mergers at high redshift (Appendix \ref{app:mergerhistories}). Wet mergers can trigger stronger starbursts and hence, increase more significantly the level of enrichment. At the same time, they induce stronger outflows which can flatten the metallicity gradients again. An example of the sequence of evolution in the shape of the metallicity gradients can be seen in Fig.~\ref{fig:innerrise} for LG1-4469 (upper two panels) and LG1-4337 (lower two panels). We find a difference in the lifetime of these features, which seems to depend on redshift. Galaxies at high redshit can recover their overall negative gradients faster because of higher probability of gas accretion. For these examples, the inner rise lasts about $150$ Myr.
While at lower redshift, mergers are less gas-rich and accretion from cold flows is very small.
This can be clearly seen in Fig.~\ref{fig:innerrise}, where  we display examples of this temporal evolution of inner broken metallicity profiles. Both galaxies, with initially  negative gradients,  get their  metallicity profiles flatten near the first apocenter due to the infall and accumulation of metal-poor gas, consistent with previous findings \citep[e.g.][]{rupke2010a,perez2011, sillero2017}. This is followed by a starburst ($\sim 0.16$ Gyr) that increases the central oxygen abundance by $\sim 0.2-0.3$ dex, inducing a break in the profile. \citet{Chen2023} report similar metallicity profiles in observed barred galaxies with high central SFRs \citep[see also][]{martinroy1995} and \citet{zahid2011}.

The lower mass LG1-4469 retains broken profiles for $\sim 0.3$ Gyr before SN feedback ejects significant gas from the central region, flattening the profile, as can be followed in Fig.~\ref{fig:gradientsintime}.
In contrast, the high-mass galaxy retains metal-rich gas against galactic winds \citep{peeples2011, Tumlinson17}, with inner breaks surviving about $\sim 1.5$ Gyr (Fig.~\ref{fig:gradientsintime}). Over time, breaks migrate outward (in units of \reff) and the level of enrichment at the break radius becomes higher ($\sim 0.1$ dex), reflecting inside-out chemical evolution tied to the star formation history. After $\sim 1.5$ Gyr, the break shifts to the outer regions, leaving the outskirts with weak or flat slopes, likely influenced by additional processes discussed in Sections \ref{subsec:outerrises} and \ref{subsec:outerdrops}.

\subsection{Inner drops and inverted gradients: The effect of energetic feedback}
\label{subsec:innerdrops}

\begin{figure*}
 \includegraphics[width=2\columnwidth]{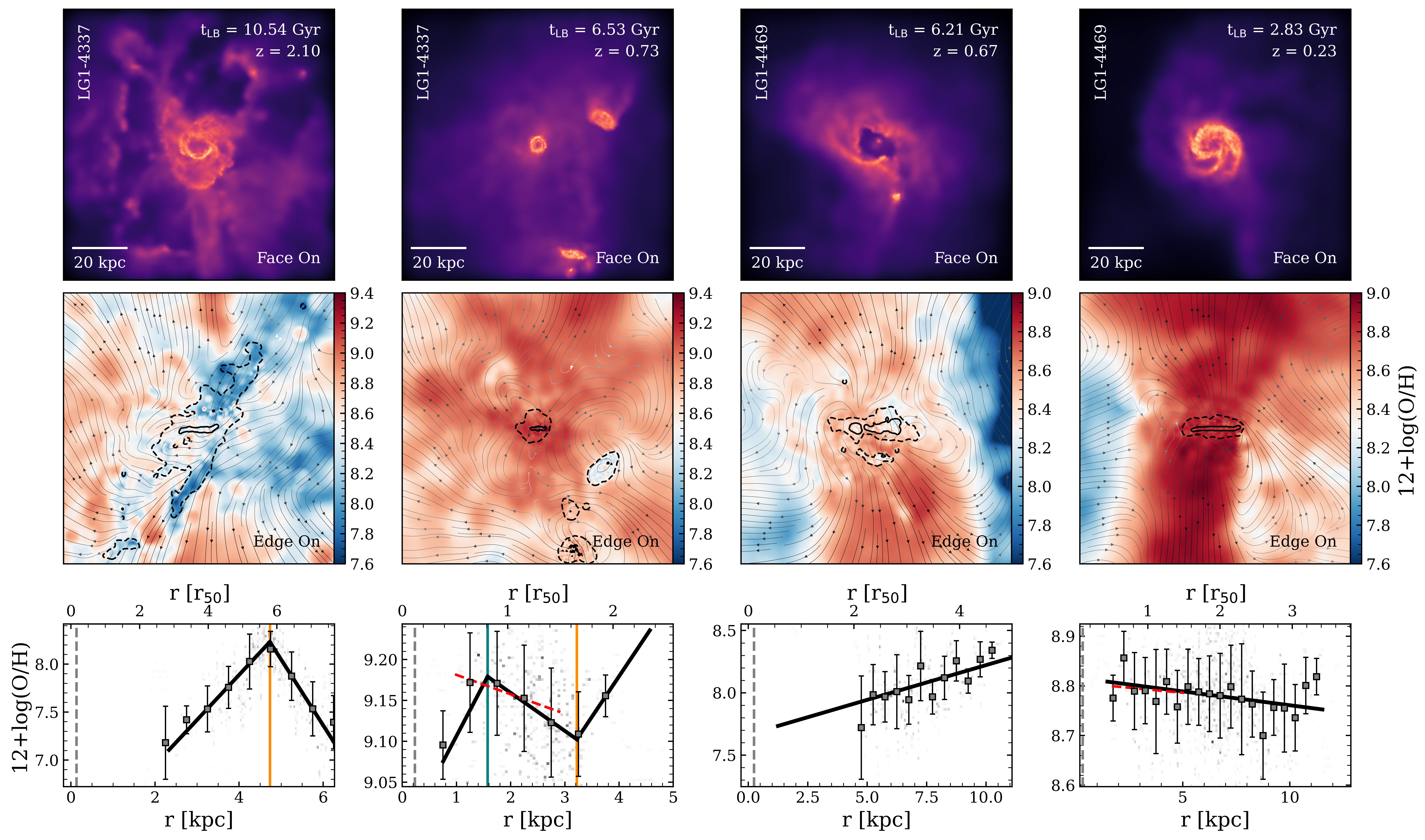}
 \caption{Examples of morphologically disrupted galaxies showing evidence of central gas ejection driven by strong energetic feedback. The top row displays face-on gas density projections; the middle row shows edge-on metallicity maps with streamlines indicating gas velocity directions; and the bottom row presents metallicity profiles, formatted as in Figs.\ref{fig:proj+prof} and \ref{fig:innerrise}. The corresponding galaxy and lookback time/redshift are labeled in the upper panels.}
 \label{fig:innerdrop}
\end{figure*}

Inner drops are identified in the metallicity profiles of the two analyzed \cielo galaxies although they are less frequent than inner rises.  Indeed, they are extensively reported in observational studies of more massive local galaxies \citep[e.g.,][]{SanchezM2018, cardoso2024}. Observational challenges have been recognized in explaining the presence of inner drops in local galaxies, highlighting that the shape of the metallicity profile may depend on the chosen diagnostic method \citep{easeman2022}. Additionally, contamination from AGN and diffuse ionized gas (DIG) can affect the identification of HII regions, impacting metallicity estimations \citep{cardoso2024}.
Hence, the comparison with observations should be always taken with caution \citep[see also][]{cornejo_subm}.

Despite these challenges, correlations between global galaxy properties and the presence of inner drops suggest a physical origin for this feature. \citetalias{SanchezM2018} find that inner drops are more common in high-mass galaxies, a result later confirmed in MaNGA galaxies by \citet{easeman2022}. Additionally, these galaxies exhibit systematically lower sSFR, suggesting that inner drops may be linked to quenching processes in the central regions. A large series of works reported lower metallicity and enhanced star formation in interacting galaxies, suggesting that tidal torques generate low metallicity inflows that trigger star formation \citep{kewley2006,ellison2020}. Recent results from \citet{garaysolis2024} found both increasing (decreasing) central abundances related with decreasing (increasing) sSFR. This is consistent with our results where we show that the metallicity evolves along the interaction and merger, modulated by the characteristics of the event and the gas available for star formation.

The lower occurrence of inner drops in the two analyzed \cielo~galaxies compared to observations could arise from the differences in simulated stellar mass range with respect to observed ones, the subgrid physics for the injection of $\alpha$-elements by SNII which is simultaneous with the injection of energy or the absence of AGN feedback, although our \cielo~sample does not include very massive galaxies which are expected to be the most affected by this feedback \citep[e.g.][]{somervilledave2015}. We highlight that numerical resolution can also affect the detection of inner features at small radii.

Interestingly, Figs. \ref{fig:ARsintime} and \ref{fig:gradientsintime} reveal that inner drops occur temporally close to strong SF bursts and inverted \midgrad, suggesting a common physical origin to both features. During these periods, strong SN feedback is expected to generate metal-loaded outflows. A visual inspection of the profiles supports this interpretation, as galaxies with inner drops are morphologically disrupted, further demonstrating the impact of energetic feedback on gas distribution. This disruption is also common in inverted \midgrad galaxies.

In Fig.~\ref{fig:innerdrop} we present a set of examples of galaxies showing different signals of interaction at different redshifts (decreasing redshift from left to right), showcasing diverse patterns in abundance distribution (bottom row), including a double break with an inner drop, two inverted \midgrad and even a negative \midgrad. All of them display ring-like distributions in the gas density maps as can be seen from this figure (top row). We also present metallicity maps from an edge-on perspective (central row), including streamlines to depict the gas motions. These maps reveal the presence of bipolar, metal-loaded outflows at all redshifts. 
If the outflows are strong enough while star formation is concentrated in outer regions of the galaxy, a clear inverted gradients can be produced as can be seen in the third column of 
Fig.~\ref{fig:innerdrop}.

Finally, the fourth column showcases a galaxy with a negative \midgrad, but with evidence of ejected gas, as seen from the lack of gas in the central region in the density projection and the metallicity profile, and proven by the strong bipolar metal-loaded outflows seen in the metallicity map. This galaxy actually corresponds to the later stage of an inner rise, previously discussed in Sec. \ref{subsec:innerrises} and shown in the right-most column of Fig. \ref{fig:innerrise}. In this case, the effect of feedback does not generate an inner drop or inverted gradient, but instead acts to erase the previously developed inner rise.

The sample of 13 inner drops identified in our case studies originate from the discussed processes at low and high redshift. In a future analyzes we expect to draw statistical results regarding an evolution of the efficiency and dependence on the characteristics of the involved physical mechanisms.

\subsection{Outer rises}
\label{subsec:outerrises}

\begin{figure*}
 \includegraphics[width=2\columnwidth]{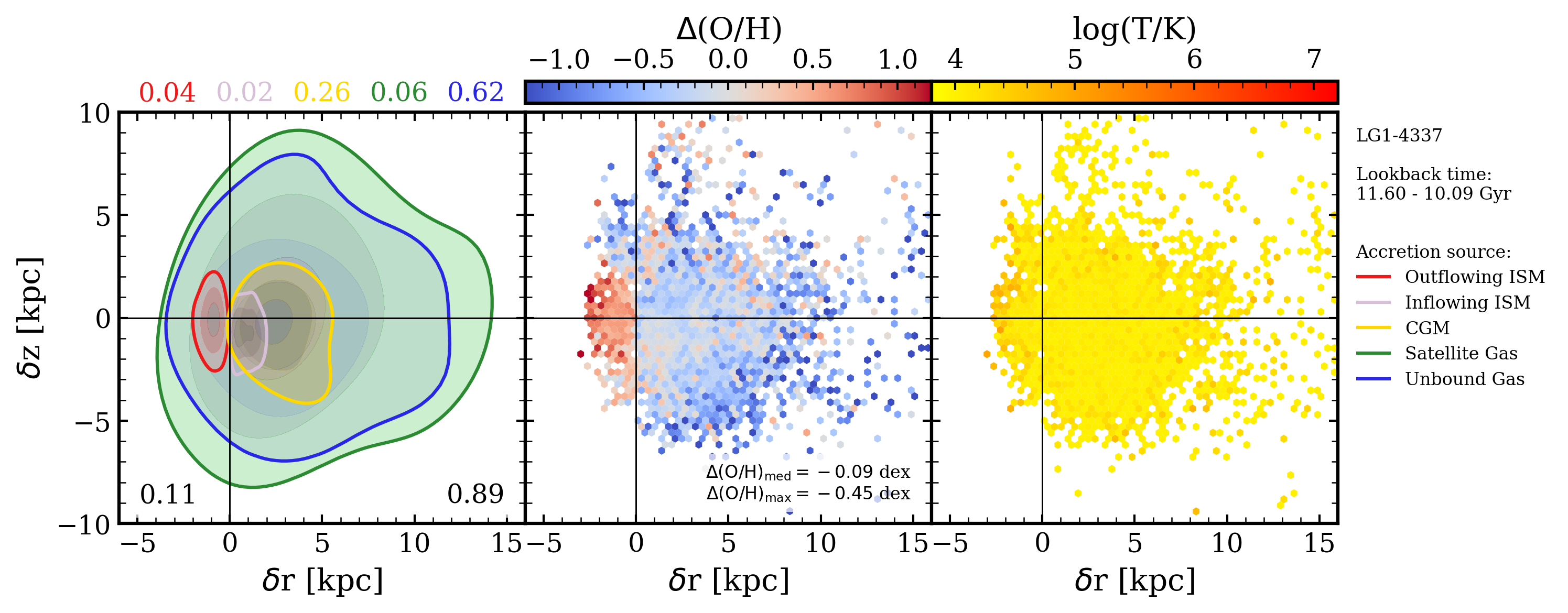}
 \includegraphics[width=2\columnwidth]{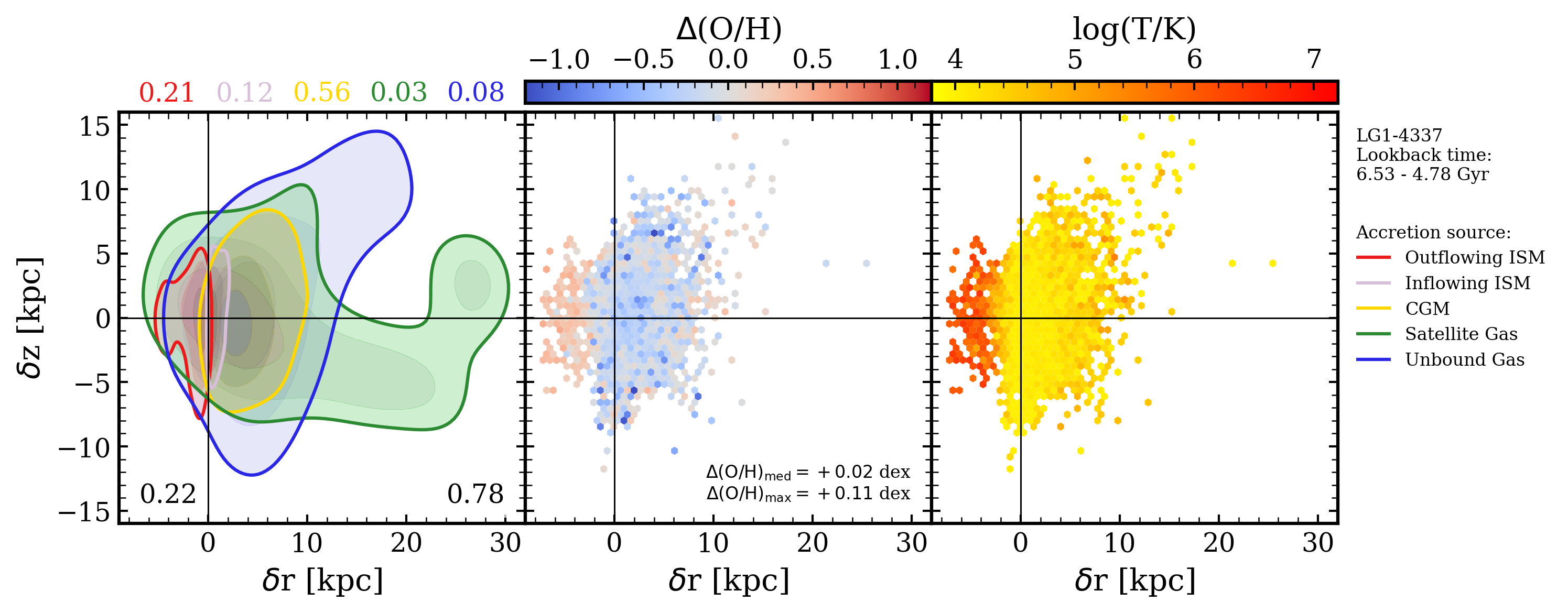}
 \caption{$\delta r - \delta z$ diagrams showing  showing the motion of the gas accreted into the outer region  of LG1-4337 from different sources (similar to Fig.~\ref{fig:drdz1}) between lookback time 11.60 Gyr and 10.1 Gyr (upper panels) and between lookback time 6.5 Gyr and 4.9 Gyr (lower panels). The diagrams in the upper panels illustrates the accretion of a high-z galaxy that exhibits an outer drop, with a dominant contribution from metal-poor unbound gas (62 per cent of the accreted gas). The diagrams in the lower panels  illustrates a low-z galaxy with an outer rise, with the clear indication recycled metal-rich gas from the CGM being the main source of accretion (56 per cent of the accreted gas).
}
 \label{fig:drdz_outer}
\end{figure*}

Observations of outer rises (also referred to as outer flattenings in the literature) have been extensively reported for galaxies in the local universe \citep{bresolin2012, Sanchez2014, SanchezM2016, SanchezM2018} and for the far side of the Milky Way \citep{minniti2020}. 

In Sec. \ref{subsec:innerrises} we discussed how an inner break, due to the continuous and extended SF activity, is able to move outwards until it is classified as an outer break. A similar scenario was proposed by \citet{wang2022_inflowmodel}, assuming a significant coplanar inflow of gas into the disc. In their model, the gradient inside the break radius is the result of the radial inflow of gas being enriched by the in-situ star formation, and the outer flattening is interpreted as a metallicity floor set by the level of enrichment of the inflow.

\citet{bresolin2012}, analysing a small sample of local galaxies, discarded the possibility that a significant fraction of the metals in the outer region are produced by in-situ star formation. Alternative possibilities, such as radial gas mixing \citep{bresolin2009, werk2011, bresolin2012} and the re-accretion of enriched material in a galactic fountain scenario driven by SN feedback \citep{dave2011, bresolin2012, Belfiore2017} has been proposed to explain these enriched outskirts. Our previous discussion on the properties of the accreted gas supports this hypothesis (Sec. \ref{subsec:accrates}). Figure \ref{fig:drdz_outer} (lower panels) showcases the motion,  $\Delta$(O/H) and temperature of the gas component in a galaxy with an outer break in the indicated time period. This figure shows that most of the accreted material that comes from the CGM is slightly more enriched than the pre-existing one, exhibiting  $\Delta {\rm (O/H)} \sim 0$, while there is also a significant part of outflowing material from the ISM, which is hot and systematically more metal-rich. Both processes coexist and simultaneously affect the properties of the ISM, resulting in a slightly flatter slope in the outer region.

\begin{figure*}
 \includegraphics[width=2\columnwidth]{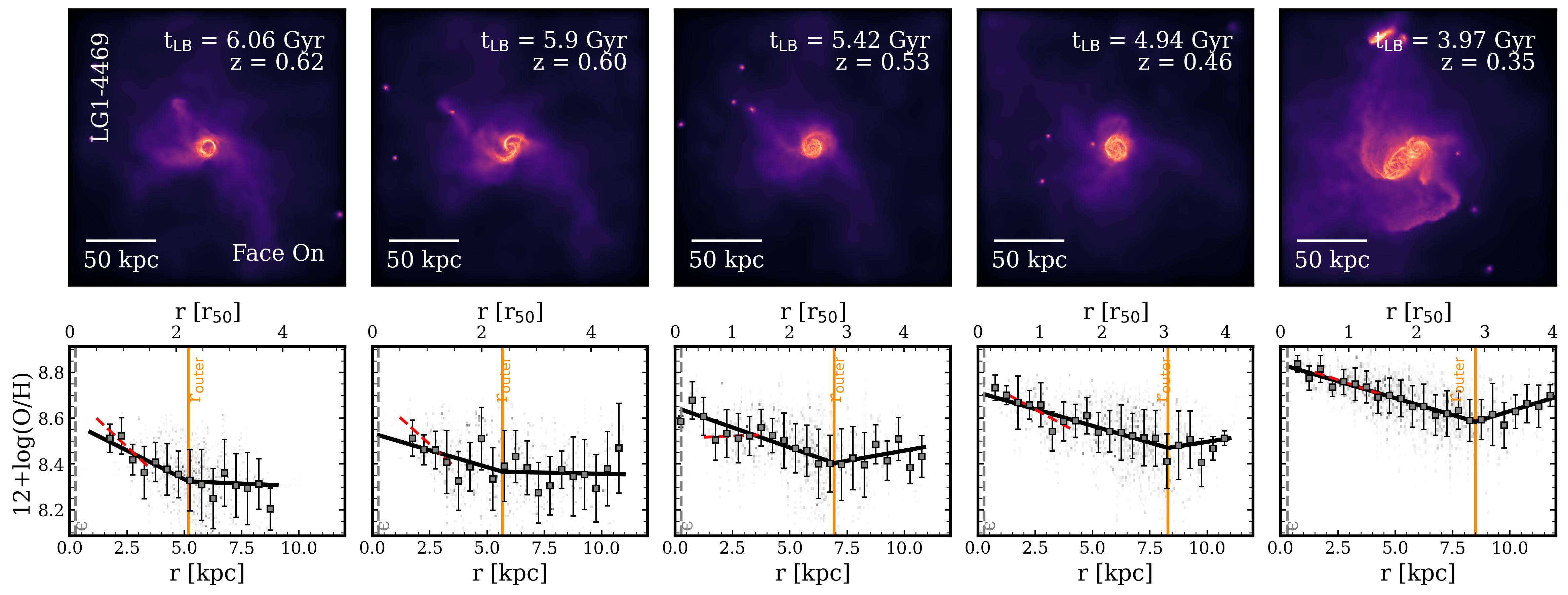}
 \includegraphics[width=2\columnwidth]{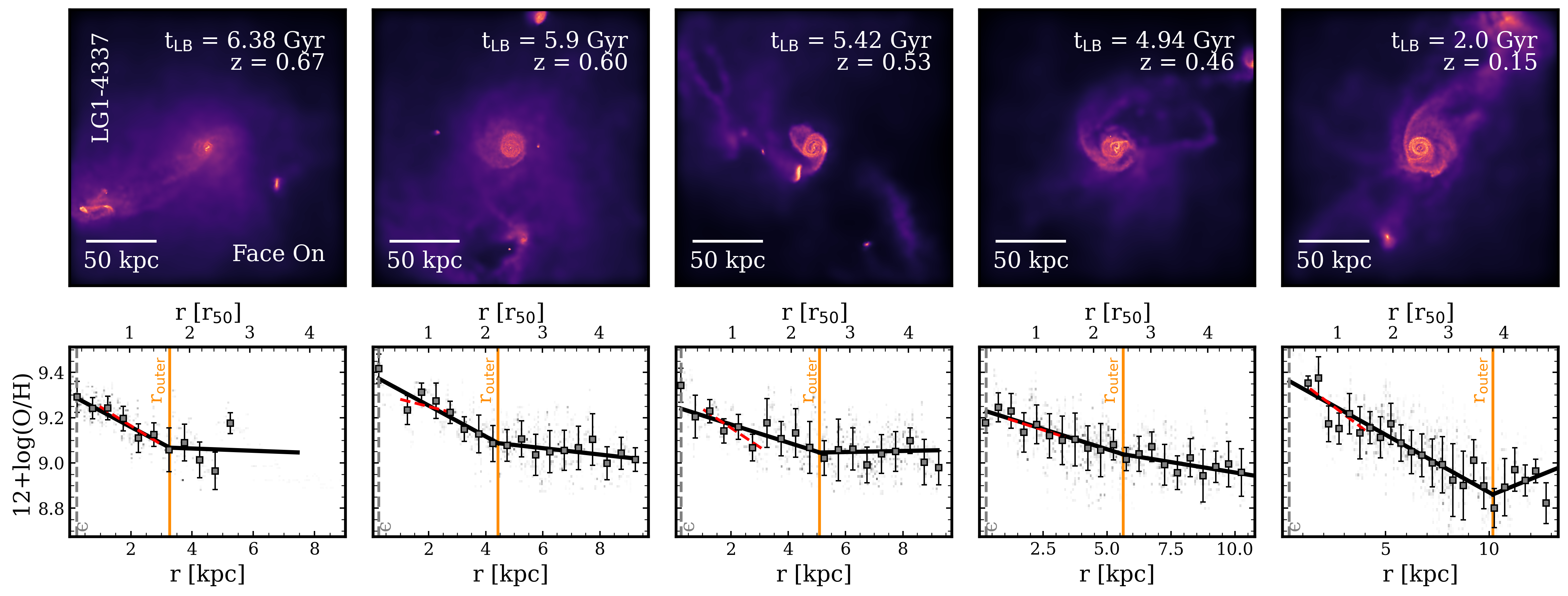}
 \caption{Similar to Fig.~\ref{fig:innerrise}, but illustrating the evolution of two examples of outer rise metallicity profiles in LG1-4469 (upper panels) and LG1-4337 (lower panels).}
 \label{fig:outerrise}
\end{figure*}

Figure~\ref{fig:outerrise} shows examples of outer breaks for both case studies (upper panels: LG1-4469; lower panels: LG1-4337). Columns one to four correspond to temporal sequences where the outer break is persistent. It can be noted that during these periods, in both cases, the abundances in the outer region are similar and the position of the break mildly evolves to larger radii. We estimated a surviving timescale for this feature that ranges within the range $1-1.5$ Gyr (see Fig. \ref{fig:gradientsintime}). The shape of the metallicity profile is also affected by galaxy interactions. As discussed in Sec. \ref{subsec:accrates}, close interactions with gas-rich satellites cause the direct accretion of gas from the companion. We found that, for LG1-4469, the sequence of outer rises shown in Fig. \ref{fig:outerrise} is disturbed due to the interaction with a gas rich satellite (lookback time $\sim 4$ Gyr; see Fig. \ref{fig:gradientsintime} and \ref{appfig:mergers4469}), whose accreted material (Fig. \ref{fig:AR_fractions}) dilutes the metallicity in the outskirts and the profile recovers the monotonic decreasing behavior. On the other hand, during the sequence shown in Fig.~\ref{fig:outerrise}, LG1-4337 has a merger event and  interactions with small satellites, most of them with small or null gas fractions. The accreted material from these companions is not enough to dilute the metallicity in the outer region, and the outer rise persists. It is also likely that the torques exerted by these satellites open up the spiral arms and favor gas mixing \citep{sillero2017}. The cold counterpart in the outflowing accreted gas seen in lower panels of Fig. \ref{fig:drdz_outer} supports this scenario.

\begin{figure*}
 \includegraphics[width=2\columnwidth]{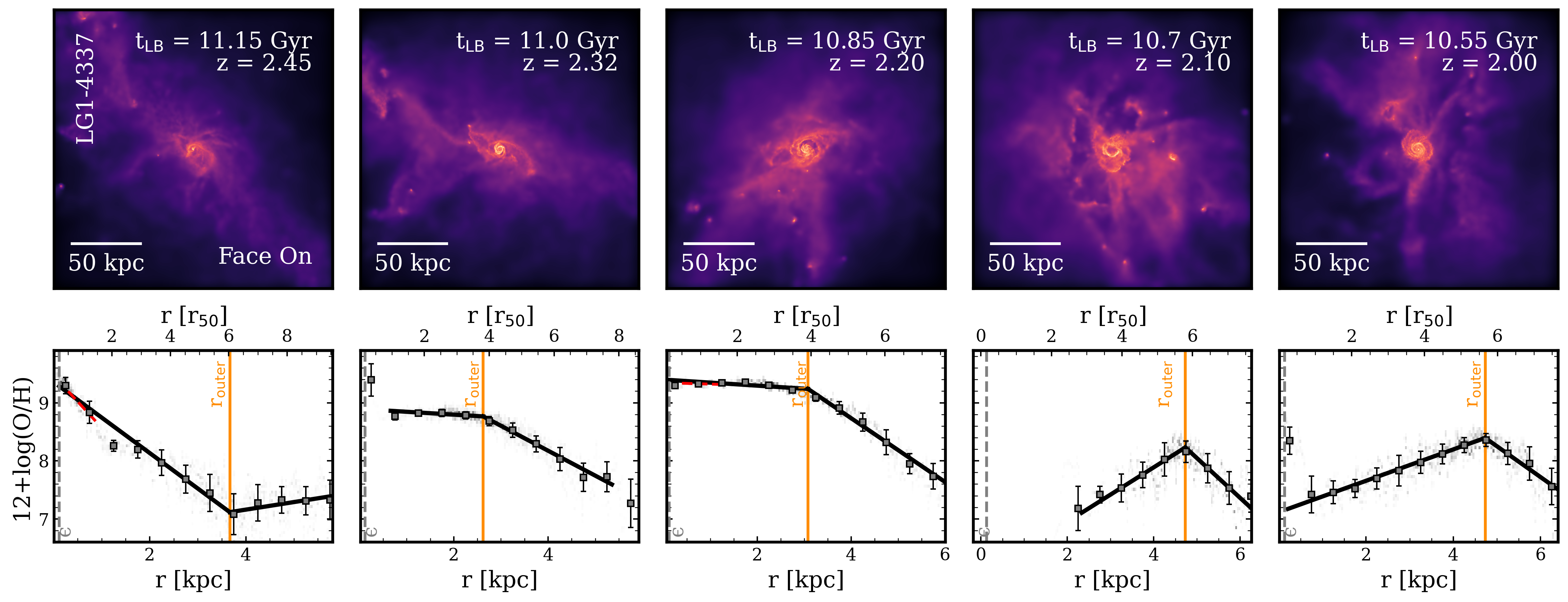}
 \caption{Similar to Fig.~\ref{fig:innerrise}, but illustrating the evolution of outer drop metallicity profiles identified in  LG1-4337.}
 \label{fig:outerdrop}
\end{figure*}

\subsection{Outer drops}
\label{subsec:outerdrops}

In the target galaxies  most outer drops occur at high redshift, where the variety and complexity of physical processes involved in disc assembly make their study particularly challenging. These profiles are typically short-lived. In Fig.~\ref{fig:outerdrop}, we present a sequence illustrating an outer drop that persists during a merger. For this galaxy, initially, the metallicity profile does not exhibit an outer drop, characterized instead by a flat  slope in the outer region, with very low metallicities. Intense star formation raises the metallicity in the mid-region while the outer region accretes low-metallicity material, generating an outer drop, a process sustained for approximately 0.3 Gyr.

The merger induces a strong, centrally concentrated starburst (see Fig. \ref{fig:ARsintime_norm}), ejecting most of the gas in the mid-region (fourth column). Despite the dispersion of ejected gas and the inversion of the gradient in the mid-region, the continuous accretion of metal-poor material maintains the outer drop (fourth and fifth columns). The metallicity map in the leftmost column of Fig. \ref{fig:innerdrop} displays the actual shape and pristine gas content of the filamentary structure involved in this sequence.

Form the upper panels of Fig.~\ref{fig:drdz_outer} we can see the characteristics of accreted material during a period when an outer drop is detected. In this case, most of gas comes from cold flows with low metallicity as expected. We acknowledge that these are two examples which have given us a clear visualization of different processes acting together to modulate the distribution of chemical elements in the ISM. As mentioned before, these results will be used to define a strategy to perform a statical study in a forthcoming paper.

\section{Conclusions}
We studied the azimuthally-averaged  metallicity profiles of the SF gas in 45 central galaxies of the \cielo~suite of simulations \citep{Tissera2025}. The sample of galaxies covers  a stellar masses mass range within the $M = [10^{8.5}, 10^{10.5}]$ at $z =0$,  representing sub Milky-Way galaxies in  the field, filamentary environments and Local Groups.

A new algorithm, DB-A, was developed to self-consistently fit multiple power laws to the metallicity profiles. DB-A selects the best combination of power laws by minimizing chi-squared values, which allows for a flexible assessment of metallicity gradients within various regions of galaxies. We applied this algorithm to \cielo~galaxies at $z=0 $ to study the global dependence of the gradients on galaxy properties. Additionally, the evolution of the metallicity gradients of two galaxies from a simulated LG analogues were analyzed as case studies to individualize the physical processes responsible of breaking the metallicity profiles. We stress the fact that these simulation do not include AGN feedback and hence, we have restricted the analysis to sub Milky Way  mass-size galaxies.

At $z=0$, the analysis revealed a diversity in shapes for the metallicity profiles,  whose properties can be summarized as follows,
\begin{itemize}

\item Approximately 38 per cent of the CIELO galaxies exhibit metallicity profiles that can be well-represented by a linear negative gradient within $\approx [0.5, 2.5]$ \reff. The rest displayed broken profiles, with 40 per cent having outer breaks, 13 per cent having inner breaks, and 9 per cent displaying double breaks in the metallicity profiles.
 
\item Linear negative metallicity profiles are consistent with the inside-out formation model, where central regions are enriched earlier and for a longer time. However, galaxies are affected by the environment and internal processes which can mix gas and stars shaping the slopes and redistributing chemical elements in the gas and stellar populations, modulating the metallicity profiles, in some cases inducing breaks in these profiles. 

\item The mid-region gradients (\midgrad) obtained by the DB-A method are comparable to the gradients obtained with a traditional single linear fit within the range [0.5, 1.5] \reff. 

\item The \midgrad, \innergrad and \outergrad agree with observational values and their relation with the stellar mass and the SFR are also consistent with observations. Smaller galaxies with lower masses tend to have flatter metallicity gradients, likely due to the influence of supernova feedback.

\item At $z=0$ the median galactocentric position of inner breaks is at $\sim 1.2$ \reff and of the outer breaks is at $\sim 2.5$ \reff for the \cielo~galaxies. The outer break radii agree well with available observations while the inner break positions tend to be larger.

\end{itemize}

The analysis of the two case study galaxies reveals that the metallicity gradient undergoes substantial variability at high redshift, transitioning to a more gradual evolution at lower redshifts. Periods of greater variability in the metallicity gradient can be linked to satellites interactions and merger events. Our study identifies inner breaks in the metallicity profiles, which may be associated with gas inflows, star formation, and energetic feedback, while outer breaks are linked to gas accretion patterns. 

Inner breaks are identified in 26 per cent of the metallicity profiles. Inner rises, which have a steep negative slope in the inner region, are observed in 72 per cent of inner break profiles.
Inner drops, where the central metallicity is diluted, are found in 28 per cent of inner break profiles.
The existence and characteristics of inner rises and inner drops in the two case studies can be  explained as:

\begin{itemize}

\item Inner rises are often caused by the efficient triggering of star formation after gas inflows toward the center coming from the ISM at lower redshifts, or from unbound gas at higher redshifts, take places. As a consequence the  central metallicity increases.

\item Inner drops can be linked to strong feedback, which causes metal-rich gas to be ejected from the center of the galaxy, evidenced by the lack of gas in the central region and disrupted gas distributions. The same physical processes can produce inverted metallicity gradients under similar conditions. At least, for the two histories of the evolutionary histories analysed in this work, inner drops seem to have shorter lifetimes than inner rises.
\end{itemize}

Outer breaks are found in 43 per cent of the metallicity profiles, characterized by a change in slope in the outer regions of the galaxy. Outer rises, where the metallicity in the outskirts is higher, are found in 68 per cent  of outer break profiles. Outer drops, where the metallicity decreases more steeply in the outer region, occur in the remaining 32 per cent, and are more common at high redshift. 
They can be explained as,
\begin{itemize}
\item Outer rises can be formed by the re-accretion of enriched material through galactic fountains, or through continuous and extended star formation in the outer regions, after a break has moved from inner regions. Enhanced gas mixing due to the tidal torques exerted in galaxy-galaxy interactions can also contribute to outer rises. 

\item Outer drops are often caused by accretion of metal-poor gas from cold flows, especially at high redshift. Mergers and interactions can also cause outer drops by enhancing star formation in the mid-region and direct accretion of low-metallicity gas in the outer region.

\end{itemize}

Double breaks exhibit both inner and outer breaks. They represent the interplay and combination of the processes that cause the individual breaks in each case.

Our findings underscore the complex interplay between internal and external processes shaping the metallicity profiles of galaxies. The detailed analysis highlights the role of gas inflows, feedback, and accretion in sculpting the distinctive features observed in metallicity gradients over cosmic time.
{\bf Our} results are consistent with previous simulations studying mergers, interactions, the CGM, and cold flows. However, our work advances the field by isolating the specific contributions of these processes during different stages of galaxy formation and evolution in a cosmological context, and, hence, underscore their potential as diagnostic tools for tracing the effects of various physical processes. Future research will involve statistical analyzes to quantify the relative impact of each mechanism, extending the analysis of the  temporal evolution to a larger sample and further understating  their role in shaping galaxy metallicity profiles

\begin{acknowledgements}
    We thank the anonymous referee for his/her constructive report which helped to improve this paper. BTC gratefully acknowledges funding by ANID (Beca Doctorado Nacional, Folio 21232155). PBT acknowledges partial funding by Fondecyt-ANID 1240465/2024 and Núcleo Milenio ERIS NCN2021\_017. This project has received funding from the European Union Horizon 2020 Research and Innovation Programme under Marie Skłodowska-Curie Actions (MSCA) grant agreement No. 101086388-LACEGAL. We acknowledge partial support by ANID BASAL project FB210003. This research was supported by the Munich Institute for Astro-, Particle and BioPhysics (MIAPbP) which is funded by the Deutsche Forschungsgemeinschaft (DFG, German Research Foundation) under Germany´s Excellence Strategy – EXC-2094 – 390783311. ES acknowledges funding by Fondecyt-ANID Postdoctoral 2024 Project N°3240644 and thanks the Núcleo Milenio ERIS. NP acknowledges support from Préstamo BID PICT Raices 2023 Nº 0002. RDT thanks the Ministerio de Ciencia e Innovación (Spain) for financial support under project grant PID2021-122603NB-C21.\\ \\

    This project has made use of: Astropy \citep{astropy:2013, astropy:2018}, Matplotlib \citep{Hunter:2007} and Py-SPHViewer \citep{Benitez_SPHviewer}. We ackonwledge the developers for their work.
\end{acknowledgements}

%

\bibliographystyle{aa} 
\bibliography{biblio}  


\begin{appendix}
\onecolumn

\section{Physical properties of \cielo~galaxies at $z=0$.}
{\renewcommand{\arraystretch}{1.1}
\begin{table*}[h!]
\caption{Physical properties and profile characteristics of \cielo~galaxies at $z=0$.}
    \centering
        \begin{tabular}{lrrrlrrrrr}
        \hline\hline
        Galaxy & $\log({\rm M_\star}/{\rm M}_\odot)$ & r$_{50}$ [kpc] & log(sSFR/yr$^{-1}$) & Profile Type & \multicolumn{3}{c}{\grad [dex r$_{50}^{-1}$]} &   \multicolumn{2}{c}{r$_{\rm br}$ [r$_{50}$]}\\
        \cmidrule(rl){6-8} \cmidrule(lr){9-10}
               &            &           &           &              &  Inner   & Middle & Outer    &   Inner & Outer \\
        \hline
        P7-2389         &  10.734 &  3.194 &  -10.969 &   Outer Break   &      --- & $-0.266$ & $-0.062$ &      --- & $ 1.713 $ \\
        LG1-4337        &  10.711 &  2.863 &  -11.720 &   Outer Break   &      --- & $-0.184$ & $-0.067$ &      --- & $ 2.785 $ \\
        P7-7805         &  10.612 &  3.398 &  -10.884 &   Inner Break   & $-0.369$ & $ 0.012$ &      --- &0.924&      --- \\
        P4-0349	        &  10.263 &  2.115 &  -10.213 &   Linear        &      --- & $-0.105$ &      --- & --- &      --- \\
        P3-0232	        &  10.216 &  2.372 &  -11.346 &   Double Break  & $-0.207$ & $ 0.171$ & $-0.241$ &0.982& $ 1.704 $ \\
        P7-8958		    &  10.169 &  2.190 &  -11.935 &   Outer Break   & ---      & $-0.269$ & $ 0.124$ & --- & 3.746\\
        P4-0000	        &  10.119 &  1.540 &  -11.002 &   Linear        & ---      & $-0.044$ & ---      & --- & --- \\
        LG2-0212        &  10.115 &  4.158 &  -10.365 &   Inner Break   & $0.006$  & $-0.245$ & ---      &1.373& --- \\
        P4-0428         &  10.092 &  2.183 &  -10.902 &   Outer Break   & ---      & $-0.197$ & $-0.316$ & --- & 2.843\\
        LG2-0220        &  9.969  &  7.168 &   -9.994 &   Double Break  & $-0.426$ & $-0.008$ & $-0.325$ &0.875& 1.782\\
        LG1-4469        &  9.906  &  3.823 &  -10.096 &   Double Break  & $-0.039$ & $-0.167$ & $ 0.004$ &0.997& 2.235\\
        P2-0214	        &  9.858  &  3.375 &  -11.526 &   Linear        & ---      & $-0.131$ & ---      & --- & --- \\
        P4-1248         &  9.767  &  3.241 &  -11.217 &   Linear        & ---      & $-0.051$ & ---      & --- & --- \\
        P7-0000		    &  9.730  &  2.175 &  -10.624 &   Linear        & ---      & $-0.010$ & ---      & --- & --- \\
        LG1-2097	    &  9.671  &  2.001 &  -11.099 &   Inner Break   & $-0.220$ & $-0.040$ & ---      &1.190& --- \\
        P7-2627		    &  9.671  &  3.198 &  -10.464 &   Linear        & ---      & $-0.061$ & ---      & --- & --- \\
        LG1-0000        &  9.662  &  3.967 &  -10.239 &   Outer Break   & ---      & $-0.281$ & $-0.172$ & --- & 2.709\\
        P4-1252         &  9.650  &  5.082 &  -10.510 &   Inner Break   & $-0.075$ & $-0.283$ & ---      &1.280& --- \\
        LG1-0053	    &  9.639  &  3.492 &  -10.396 &   Linear        & ---      & $-0.131$ & ---      & --- & --- \\
        LG1-2250	    &  9.526  &  2.968 &  -10.893 &   Outer Break   & ---      & $-0.209$ & $-0.018$ & --- & 2.182\\
        P3-0271	        &  9.517  &  3.644 &  -10.859 &   Inner Break   & $-0.106$ & $-0.664$ & ---      & 0.945 & --- \\
        P4-1266	        &  9.494  &  3.757 &  -10.213 &   Linear        & ---      & $-0.265$ & ---      & --- & --- \\
        LG1-2208     	&  9.399  &  1.965 &  -11.183 &   Linear        & ---      & $-0.061$ & ---      & --- & --- \\
        P4-1258         &  9.393  &  3.305 &  -10.493 &   Outer Break   & ---      & $-0.077$ & $-0.393$ & --- & 2.951\\
        P7-2696         &  9.362  &  0.967 &  -11.629 &   Outer Break   & ---      & $-0.130$ & $-0.033$ & --- & 3.023\\
        P3-0000         &  9.284  &  4.749 &  -10.554 &   Linear        & ---      & $-0.114$ & ---      & --- & --- \\
        P4-0018         &  9.279  & 1.493  &  -11.097 &  Outer Break    & ---      & $-0.235$ & $-0.463$ & --- & 3.751\\
        LG1-0105        &  9.255  & 2.473  &  -10.805 &  Outer Break    & ---      & $-0.280$ & $-0.090$ & --- & 1.668\\
        LG1-0032        &  9.224  & 2.789  &   -9.258 &  Outer Break    & ---      & $-0.377$ & $ 0.033$ & --- & 2.655\\
        LG1-2298        &  9.214  & 2.678  &  -11.307 &  Linear         & ---      & $-0.129$ & ---      & --- & --- \\
        P2-0000         &  9.177  & 4.144  &  -10.005 &  Double Break   & $0.022$  & $-0.403$ & $ 0.495$ &1.170& 2.577\\
        P7-2717         &  9.099  & 0.955  &  -11.034 &  Linear         & ---      & $-0.154$ & ---      & --- & --- \\
        LG1-0107        &  9.056  & 3.083  &  -10.186 &  Outer Break    & ---      & $-0.390$ & $ 0.569$ & --- & 2.743\\
        LG1-0081        &  9.055  & 1.615  &   -9.765 &  Linear         & ---      & $-0.219$ & ---      & --- & --- \\
        LG1-4647        &  9.018  & 1.694  &  -10.259 &  Inner Break    & $-0.467$ & $-0.093$ & ---      &1.496& --- \\
        P3-0298         &  9.011  & 3.031  &   -9.872 & Outer Break     & ---      & $-0.417$ & $ 0.324$ & --- & 1.767\\
        P7-0181         &  8.917  & 2.541  &  -10.825 & Linear          & ---      & $-0.120$ & ---      & --- & --- \\
        P7-0192         &  8.808  & 1.030  &  -11.335 & Linear          & ---      & $-0.046$ & ---      & --- & --- \\
        P7-2727         &  8.789  & 2.601  &  -10.159 & Outer Break     & ---      & $ 0.132$ & $-0.565$ & --- & 1.736\\
        P7-2774         &  8.392  & 1.148  &  -10.035 & Linear          & ---      & $-0.110$ & ---      & --- & --- \\
        P7-9110         &  8.242  & 1.060  &  -10.014 & Outer Break     & ---      & $-0.119$ & $ 0.282$ & --- & 2.942\\
        P7-2736         &  8.172  & 0.967  &  -10.165 & Outer Break     & ---      & $-0.141$ & $-0.029$ & --- & 1.694\\
        P7-2763         &  8.105  & 1.053  &   -9.847 & Outer Break     & ---      & $-0.145$ & $ 0.379$ & --- & 3.739\\
        P7-2780         &  8.043  & 1.065  &  -10.121 & Outer Break     & ---      & $-0.283$ & $-0.086$ & --- & 2.279\\
        P7-0258         &  8.025  & 1.647  &   -9.390 & Linear          & ---      & $-0.064$ & ---      & --- & --- \\
        \hline
        Median                    & & & & & -0.157 & -0.131 & -0.048 & 1.083 & 2.616 \\
        25$^{\rm th}$ p.  & & & & & -0.332 & -0.265 & -0.224 & 0.955 & 1.771 \\
        75$^{\rm th}$ p.  & & & & & -0.048 & -0.064 &  0.102 & 1.257 & 2.917 \\
        \hline
        \end{tabular}
    \tablefoot{Slopes in the inner and outer regions are displayed when breaks in the abundance distribution are found. The break radius, normalized by \reff, is also included.}
    \label{tab:tab_z0}
\end{table*}}
\newpage
\section{Merger histories}
\label{app:mergerhistories}

\begin{figure}[h!]
 \includegraphics[width=0.49\columnwidth]{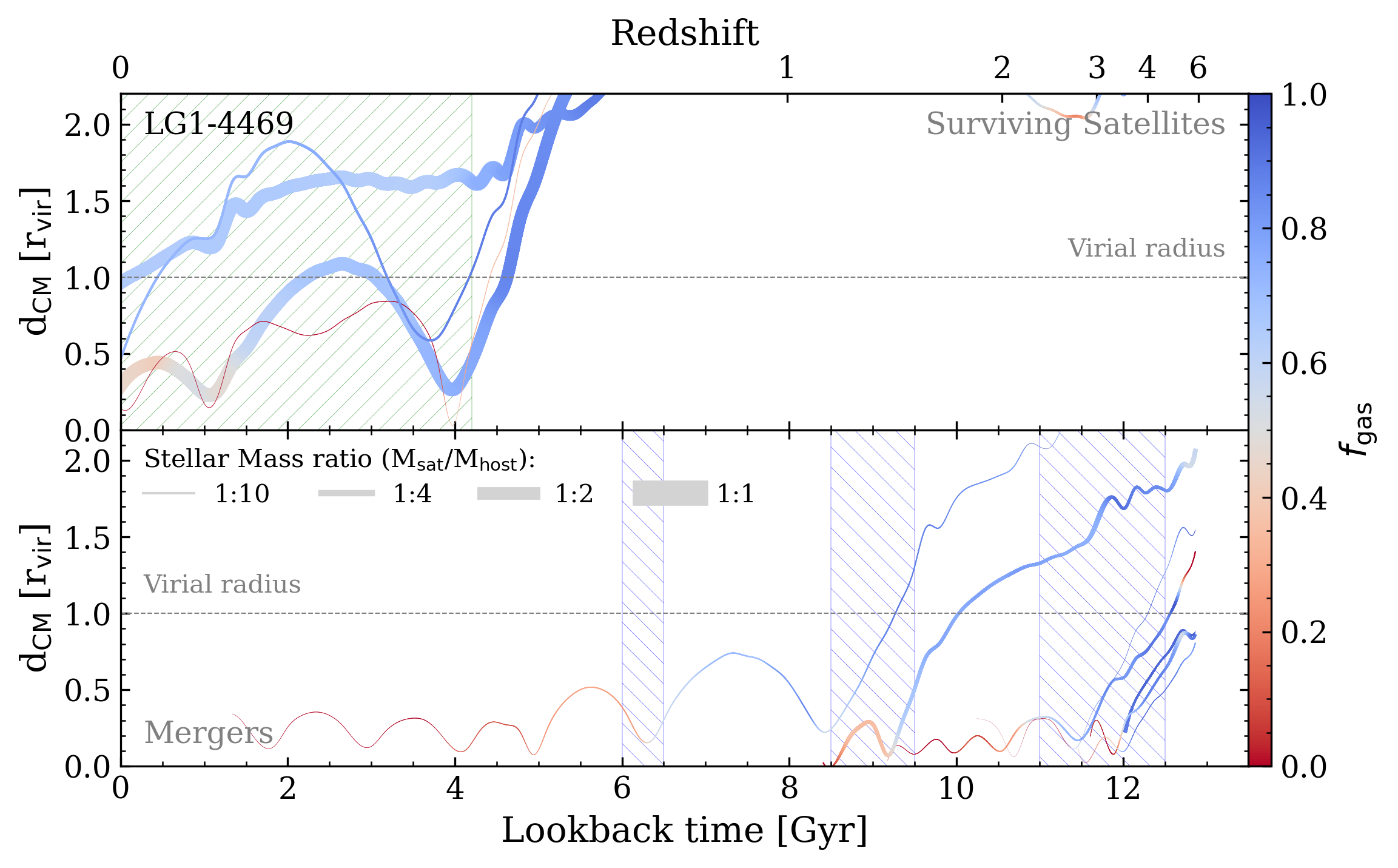}
 \includegraphics[width=0.49\columnwidth]{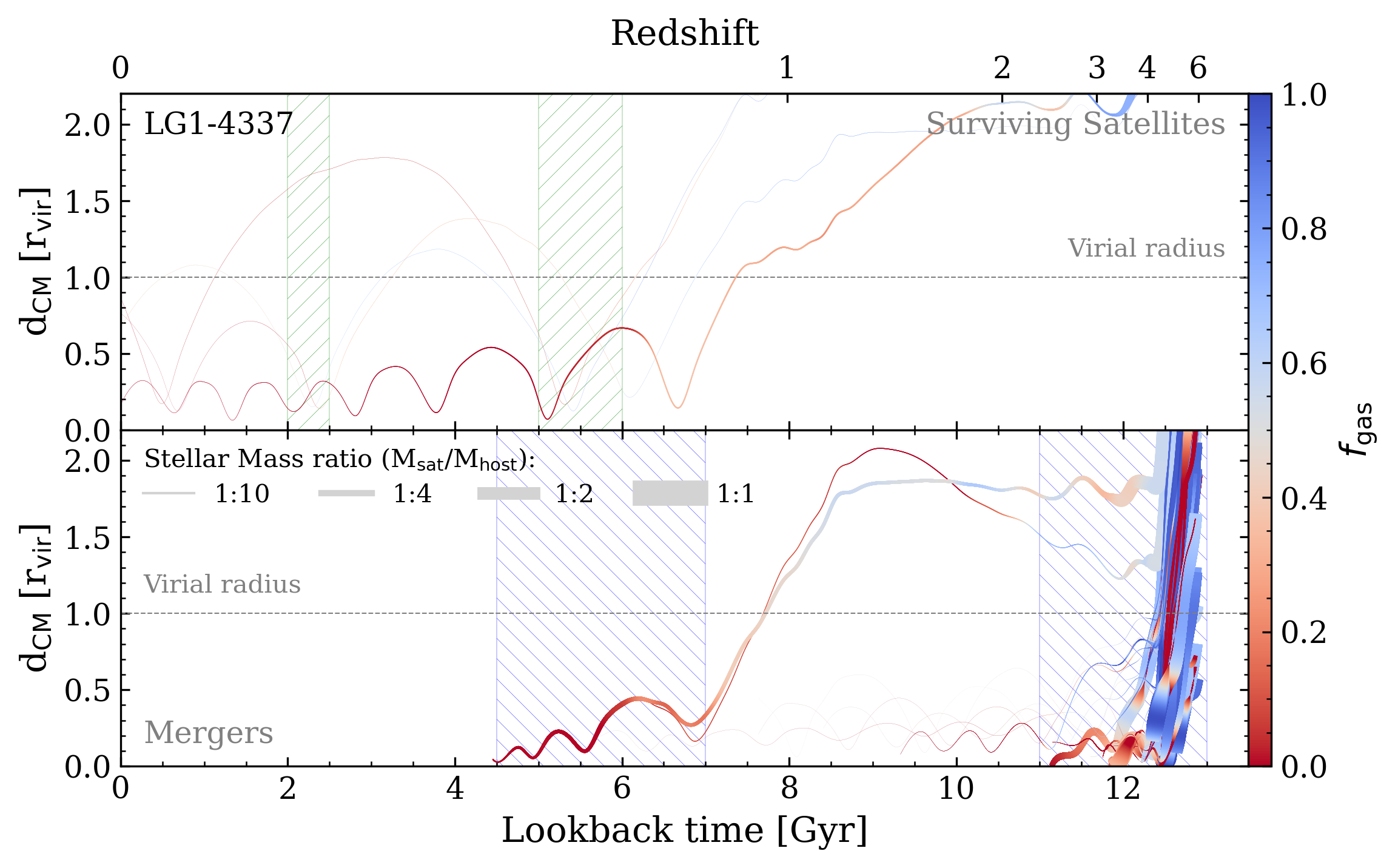}
 \caption{Time evolution of the physical separation between the center of mass of the central galaxy (left panel: LG1-4469; right panel: LG1-4337) and its surviving satellites (upper panel) and merged galaxies (lower panel). Each continuous line represents a different satellite or merged galaxy, with line thickness proportional to their stellar mass ratio. The line color indicates the gas fraction of the satellite/merged galaxy at different stages of the interaction. Key events are highlighted with green and blue shaded regions, consistent with those shown in Fig. \ref{fig:gradientsintime}.}
 \label{appfig:mergers4469}
\end{figure}

In Sections \ref{subsec:satellites} and \ref{subsec:mergers}, we discussed how interactions between the central galaxies and their satellites or mergers were traced back in time using their merger trees to identify key events contributing to their assembly histories. For illustration purposes we selected specific periods of either mergers or interactions that clearly highlight the effect of these physical processes on the metallicity distribution. Figure \ref{appfig:mergers4469} provide a detailed view of these interactions, with the upper panels depicting surviving satellites and the lower panels showing mergers. These interactions are tracked through the physical separation of their centers of mass, along with the evolution of their stellar mass ratios and gas fractions at various stages. The most significant events, based on proximity, stellar mass ratios, gas fractions, and coupled interactions, are highlighted in green and blue shaded regions.

The assembly histories of the two galaxies differ markedly. Wet mergers are frequent at high redshift for both, but LG1-4337, the more massive galaxy, experiences a greater number of massive mergers and completes its assembly earlier. At lower redshift, merger galaxies are typically 25 per cent the mass of the central galaxy, with higher gas fractions observed in LG1-4469. This trend also applies to their surviving satellites. Note that LG1-4469 is undergoing a major interaction with a gas-rich companion at $z=0$. Coupled interactions, in both merging galaxies and surviving satellites, are common in the evolution of both systems \citep[see, e.g.,][]{gamez2024}.

\newpage
\section{Normalized $\dot{\mathcal{M}}$ and SFR}
\label{appendix1}

\begin{figure}[h!]
 \includegraphics[width=1\columnwidth]{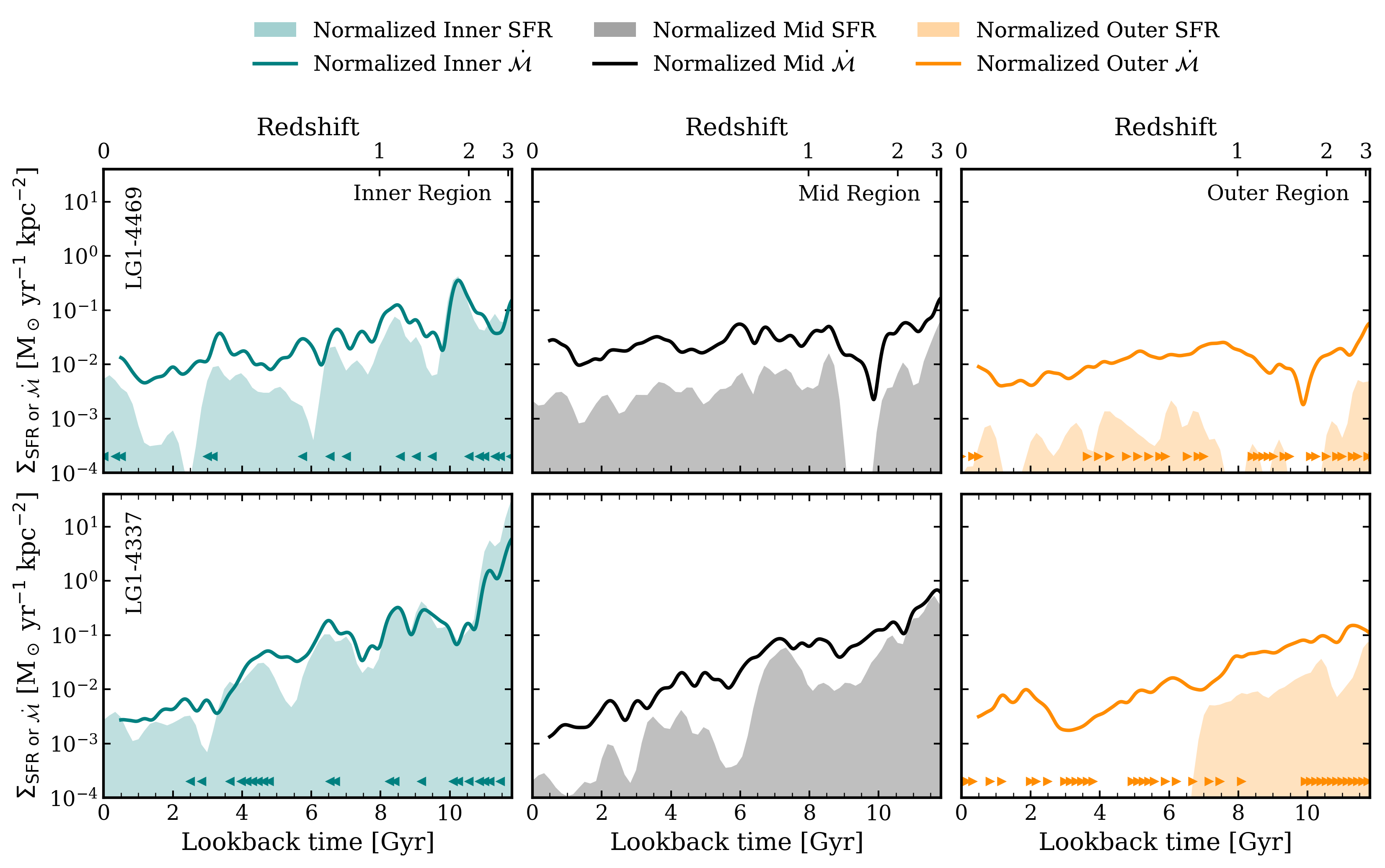}
 \caption{Same as Fig. \ref{fig:ARsintime} but normalised by the projected area of each region.}
 \label{fig:ARsintime_norm}
\end{figure}

In Fig. \ref{fig:ARsintime} we show the mass of accreted gas ($\dot{\mathcal{M}}$) and stars formed (SFR) per unit time. However, the inner, mid and outer regions will naturally have different sizes, and they will evolve in time as well. In Fig. \ref{fig:ARsintime_norm} we display the ($\dot{\mathcal{M}}$) and SFR normalised by the projected area of each region. The normalization reveals that the amount of accreted gas and stars formed per unit area is highest in the inner region, even though it tends to be the smallest in size. As discussed in Section \ref{subsec:accrates}, the normalized values also show a strong correlation between SFR and accreted gas in the inner region, while the mid and outer regions show an offset between these quantities.

\end{appendix}
\end{document}